\documentclass[times,authoryear,3p]{elsarticle}
\journal{Journal XY}
\usepackage[labelfont=bf]{caption}

\usepackage{amsmath,amsfonts,amssymb,amsthm,color,epsfig,graphicx,hyperref,url}

\usepackage{natbib}
\usepackage{dsfont}
\usepackage{upgreek, mathrsfs}
\usepackage{tikz}

\usepackage{rotating}
\usepackage{array} %alignment resp graphics in table
\usepackage{longtable}
\usepackage{enumitem}
\usepackage{calc} 
\usepackage{booktabs,caption}
\DeclareCaptionLabelFormat{continued}{#1~#2 (cont.)}
\newcommand{\bbz}{\mathbb{Z}}
\newcommand{\bbn}{\mathbb{N}}

\newcommand{\bbe}{\mathbb{E}}
\newcommand{\var}{\mathbb{V}}
\newcommand{\bbp}{\mathbb{P}}
\newcommand{\cov}{\text{Cov}}
\newcommand{\corr}{\text{Corr}}

\newcommand{\nb}{\text{NB}}
\newcommand{\poi}{\text{Poi}}
\newcommand{\bin}{\text{Bin}}

\newcommand{\indicator}[1]{\mathds{1}_{\{ #1 \}}}
\newcommand{\pgf}{\operatorname{pgf}}

\newcommand{\bftheta}{\boldsymbol{\theta}}
\newcommand{\bfvartheta}{\boldsymbol{\vartheta}}

\newcommand{\brackets}[1]{\left( #1 \right)}

\theoremstyle{plain}% Theorem-like structures provided by amsthm.sty
\newtheorem{theorem}{Theorem}[section]

\newtheorem{proposition}[theorem]{Proposition}

\newtheorem{corollary}[theorem]{Corollary}

\theoremstyle{definition}
\newtheorem{definition}[theorem]{Definition}

\newtheorem{remark}[theorem]{Remark}
\newtheorem{example}[theorem]{Example}

\begin{document}

\begin{frontmatter}

\title{A Class of Higher-Order INAR Random Fields for Poisson Counts and Beyond}

\author[hsu]{Christian H.\ Wei\ss\corref{cor}}
\ead{weissc@hsu-hh.de}

\author[hsu]{Angelika Silbernagel\fnref{fn1}}
\ead{silbernagel@hsu-hh.de}

\cortext[cor]{Corresponding author. ORCID: \href{https://orcid.org/0000-0001-8739-6631}{\nolinkurl{0000-0001-8739-6631}}.}
\fntext[fn1]{ORCID: \href{https://orcid.org/0009-0002-6993-244X}{\nolinkurl{0009-0002-6993-244X}}.}
\address[hsu]{Department of Mathematics and Statistics, Helmut Schmidt University, Holstenhofweg 85, 22043 Hamburg, Germany.}

\begin{abstract}
Existing integer-valued autoregressive (INAR) models for count random fields suffer from difficulties in characterizing the stationary marginal distribution and in computing conditional probabilities (as required for likelihood inference). To overcome these drawbacks, the novel class of combined INAR (CINAR) models is proposed, which both exhibits the classical autoregressive dependence structure and allows to specify the marginal distribution within the wide class of discrete self-decomposable distributions. In particular, CINAR random fields can be equipped with a Poisson or negative-binomial marginal distribution. The CINAR's key stochastic properties are derived (including a simple expression for conditional probabilities), and special cases as well as possible extensions are discussed. Approaches for parameter estimation are developed and investigated, and the practical relevance of the novel CINAR family is demonstrated by an agricultural data application. 
\end{abstract}


\end{frontmatter}

\section{Introduction}
\label{section: Introduction}
In many applications, we are concerned with data that are located on a regular two-dimensional grid; the corresponding data-generating process (DGP) is commonly referred to as a random field. Such grid data and random fields can be understood as ``plane extensions'' of time series and stochastic processes, respectively. Hence, it is not surprising that many models for random fields arise from analogous time-series models. In the case of continuously distributed random fields, for example, the random-field versions of the classical autoregressive (AR) and moving-average (MA) models as developed by \citet{whittle54, besag74, haining78, basu93} are often used in applications. 
Here, one distinguishes between unilateral and multilateral model structures, where the observations are affected by only one particular direction or all directions, respectively \citep{whittle54}. While a multilateral structure is sometimes more plausible in applications, it goes along with difficulties in statistical inference (e.g., with respect to likelihood computations, see \citealp{tjostheim83}). Therefore, (approximate) unilateral structures are often preferred in practice. 
Moreover, they are the natural choice if there is a natural ordering to the sites, e.g., when modeling real spatial data which has a slope of the ground or wind coming from a specific direction as an underlying mechanism \citep{basu93}. Examples include the number of plants affected by fungal disease or pests which spread predominantly with prevailing winds, seeds that are washed downhill after a heavy rain, or the number of clogged drains in a stormwater drainage systems in urban areas. Another prominent application is given by migrational data. People might flee in one direction, e.g, because of war, famine or natural disasters. The random field could then be interpreted as a snapshot that explains the distribution of the population across a grid. 

\smallskip
In the present research, we are not concerned with continuously distributed random fields, but count random fields $(X_{s,t}) = (X_{s,t})_{s,t\in\bbz=\{\dots, -1, 0, 1, \dots\}}$ having $\bbn_0 = \{0,1,\ldots\}$ as their range. Corresponding applications with count grid data can be found in various fields of practice, such as  agriculture \citep{tabandeh24}, biology \citep{chutoo21}, or epidemiology \citep{yang25}. As with continuously distributed random fields, it makes sense to adapt established models from the field of count time series to the case of count random fields. In the seminal article by \citet{ghodsi12}, a spatial analog of the famous first-order integer-valued AR (INAR$(1)$) model known from count time series analysis (see \citealp{weiss18} for a comprehensive survey) was introduced, which the authors originally referred to as the ``first-order spatial INAR (SINAR$(1,1)$)'' model. In what follows, however, we use the terminology of \citet{sil_wei_26} who refer to this model as the INAR$(1,1)$ \emph{random field}, as this notion emphasizes the grid structure of the data and avoids confusion with other types of spatial count data (for the latter, see \citet{glaser17,jung22} and the references therein). The (unilateral) INAR$(1,1)$ random field is defined by the recursion
\begin{equation}
    \label{eq: model}
    X_{s,t} = \alpha_{01} \circ X_{s,t-1} + \alpha_{10} \circ X_{s-1,t} + \alpha_{11} \circ X_{s-1,t-1} + \varepsilon_{s,t},
\end{equation}
where the model's dependence parameters satisfy 
\begin{equation}
    \label{eq: parameter condition}
    \alpha_{01}, \alpha_{10}, \alpha_{11} \in [0,1)
    \quad\text{and}\quad
    \alpha_{01}+\alpha_{10}+\alpha_{11}<1,
\end{equation}
and where the count innovations $(\varepsilon_{s,t})=(\varepsilon_{s,t})_{s,t\in\bbz}$ are independent and identically distributed (i.i.d.)\ with mean $\mu_\varepsilon>0$ and variance $\sigma^2_\varepsilon>0$. Moreover, introducing the short-hand notation $\mathcal{P}\!_{s,t}=\{X_{s-k,t-l} : k\geq1 \text{ or } l\geq 1\}$ for the ``past'' corresponding to ``time'' $(s,t)$ (note the ``or'' in the definition of~$\mathcal{P}\!_{s,t}$), the innovation $\varepsilon_{s,t}$ at ``time'' $(s,t)$ is assumed to be independent of~$\mathcal{P}\!_{s,t}$ for all $s,t\in\bbz$. Finally, the operation ``$\circ$'' in \eqref{eq: model} refers to the so-called ``binomial thinning operator'', which serves as an integer substitute of the ordinary multiplication \citep{weiss08b}. For a discrete count random variable $X$ with range $\bbn_0$ and a constant $\alpha \in (0,1)$, it is defined by $\alpha \circ X := \sum^X_{i=1} Z_i$ (together with the convention that empty sums are equal to zero), where the counting series $(Z_i)_{i\in\bbn=\{1, 2, \dots\}}$ is a sequence of i.i.d.\ Bernoulli random variables with $\bbp(Z_i=1)=\alpha$, which is independent of $X$. The boundary values $\alpha\in\{0,1\}$ can be included by the conventions $0 \circ X := 0$ and $1 \circ X:=X$.
The thinnings at point $(s,t)$ appearing in \eqref{eq: model} are assumed to be performed independently of each other, to be independent of $(\varepsilon_{i,j})$, and, see \citet{sil_wei_26} for a discussion, to be independent of $\mathcal{P}\!_{s,t}$ as well.

\smallskip
The INAR$(1,1)$ random field \eqref{eq: model} of \citet{ghodsi12} has been studied and modified in many subsequent works, see \citet{ghodsi15,chutoo21,sassi23,ghodsi24,karlis24,tabandeh24,ghodsi25,yang25,sil_wei_26} for details. 
Moreover, \citet{sil_wei_26} extended the INAR$(1,1)$ model to higher-order autoregressions, namely to the INAR$(p_1,p_2)$ model of order $(p_1,p_2)\in\bbn^2$ being defined by the recursion
\begin{equation}
    \label{eq: generalized model}
    X_{s,t} = \sum_{(i,j)\in\mathcal{S}} \alpha_{ij} \circ X_{s-i,t-j} + \varepsilon_{s,t}
    \quad\text{with }
    \mathcal{S}:=\big\{(i,j) \,|\, 0 \leq i \leq p_1,\, 0 \leq j \leq p_2,\, (i,j) \neq (0,0)\big\},
\end{equation}
and derived its key stochastic properties. 
However, these ``default'' INAR models for count random fields, as defined in the aforementioned references, have the drawback that closed-form expressions for the stationary marginal distribution are difficult to obtain. This problem is caused by the multiple thinnings at each point $(s,t)$, and it is already evident in the first-order case $p_1=p_2=1$, recall \eqref{eq: model}. For example, it is not clear how spatially dependent Poisson (Poi) counts could be obtained by an INAR random field, i.e., how the innovations' distribution has to be chosen such that the generated~$X_{s,t}$ are Poi-distributed (if possible at all). This differs from the time-series case, where at least the INAR$(1)$ model can be equipped with a Poi-marginal distribution (and some further common distributions, see Section~\ref{subsection: Stochastic Properties of CINAR Random Fields} for details). Up to now, the only ARMA-like approach for Poi-grid data appears to be the integer-valued MA (INMA) random field of \citet{sil_wei_26b}, which can be equipped with a Poi-marginal distribution for any model order. On the other hand, MA-type models have a memory that ends abruptly, whereas the memory of AR-type models fades out gradually and, thus, often better matches real-world data. Therefore, it would be attractive for practice to have an INAR-type model for count random fields that allows for a Poi-marginal distribution (or other common distributions) at the same time. Such a specified type of marginal distribution could be useful for interpretations and would allow for simple checks of model adequacy.

\begin{remark}
\label{remark:gcrf}
While the focus of the present research is on ARMA-like models for count random fields, it is worth noting that also different modeling approaches have been proposed in the literature. In particular, different types of Gaussian copula random fields should be mentioned, which can also be equipped with a specified marginal distribution. However, likelihood computations for parameter estimation are quite challenging for these models, see \citet{masarotto12,kazianka13,hughes15,han20} and the references therein for a discussion. By contrast, we will recognize that the proposed CINAR random fields possess rather simple closed-form expressions for conditional probabilities such that likelihood evaluations are computationally cheap.
\end{remark}
The drawback of having an unclear marginal distribution is also well known from the INAR$(p)$ time-series model with $p>1$ according to \citet{du91} (whereas the Poi-INAR$(p)$ model of \citet{alzaid90} does not exhibit the typical AR$(p)$ autocorrelation structure). A possible solution that combines both an AR$(p)$-like  autocorrelation function (ACF) and a Poi-marginal distribution was developed by \citet{weiss08c} based on preliminary work by \citet{zhu06}. Their idea was to combine the INAR$(1)$ time-series model (involving only one thinning per time) with a random-selection mechanism, with the resulting models being referred to as \emph{combined INAR (CINAR) models}. In what follows, we adapt this idea to the case of count random fields by developing the novel class of CINAR random fields, see Section~\ref{section: The Combined INAR Random Field}. There, we also derive the key stochastic properties of the proposed (unilateral) CINAR family, discuss important special cases, and present possible future extensions. Section~\ref{section: Parameter Estimation} develops possible approaches for the parameter estimation of CINAR models, the finite-sample performances of which are investigated in Section~\ref{section: Performance of Parameter Estimation} by a comprehensive simulation study. The practical relevance of the novel CINAR class is demonstrated in Section~\ref{Data Example: Yields from an Agricultural Experiment}, where a real-world data application from an agriculutural experiment is presented. Finally, we conclude in Section~\ref{section: Conclusions and Future Research} and outline directions for future research.

\numberwithin{theorem}{subsection}

\section{The Combined INAR Random Field}
\label{section: The Combined INAR Random Field}
Inspired by the works of \citet{zhu06,weiss08c} on count time series, let us propose the class of combined INAR models for count random fields.

\begin{definition}
    \label{def: CINAR}
    Like in \eqref{eq: generalized model}, let $\mathcal{S}:=\big\{(i,j) \,|\, 0 \leq i \leq p_1,\, 0 \leq j \leq p_2,\, (i,j) \neq (0,0)\big\}$ for given $p_1,p_2\in\bbn$. We say that the random field $(X_{s,t})_{s,t\in\bbz}$ follows a (unilateral) \emph{CINAR$(p_1,p_2)$ model} if it satisfies the recursion
    \begin{equation}
        \label{eq: CINAR recursion}
        X_{s,t}
        = \sum_{(i,j)\in\mathcal{S}} D_{s,t;i,j} \cdot (\alpha \circ_{s,t} X_{s-i,t-j}) + \varepsilon_{s,t},
    \end{equation}
    where
    \begin{itemize}
        \item $(\varepsilon_{s,t})$ are i.i.d.\ count innovations and $\alpha \in (0,1)$ is a constant;
        
        \item $(\mathbf{D}_{s,t})_{s,t\in\bbz}$ is an i.i.d.\ random field of $|\mathcal{S}|$-dimensional ``decision'' random vectors $\mathbf{D}_{s,t}=(D_{s,t;0,1}, \dots, D_{s,t;p_1,p_2})$ that follow the multinomial distribution $\text{MULT}(1;\phi_{01}, \dots, \phi_{p_1p_2})$ with probabilities~$\phi_{ij}$ satisfying $\sum_{(i,j)\in\mathcal{S}} \phi_{ij} =1$ (i.e., where exactly one of the $|\mathcal{S}|$ components takes the value 1 and all others equal 0), which are also independent of the innovations $(\varepsilon_{s,t})$;
        
        \item $\varepsilon_{s,t}$ and $\mathbf{D}_{s,t}$ are independent of the ``past'' $\mathcal{P}\!_{s,t}$;
        
        \item the thinnings at ``time'' $(s,t)$ are performed independently of each other, of $(\varepsilon_{s,t})$, $(\mathbf{D}_{s,t})$, and the ``past'' $\mathcal{P}\!_{s,t}$.
    \end{itemize}
\end{definition}
Note that the subscript ``$s,t$'' at the thinnings in \eqref{eq: CINAR recursion} emphasizes that the thinnings are executed at point $(s,t)$ and are not identical to thinnings being executed at other locations. For the sake of readability, we omit this subscript whenever there is no risk of a misunderstanding. The recursion \eqref{eq: CINAR recursion} in Definition~\ref{def: CINAR} states that $X_{s,t}$ is equal to $\alpha \circ X_{s-i,t-j}+\varepsilon_{s,t}$ with probability $\phi_{i,j}$, for $(i,j)\in\mathcal{S}$. Hence, by contrast to the ordinary INAR$(p_1,p_2)$ recursion \eqref{eq: generalized model}, the CINAR$(p_1,p_2)$ random field uses only one thinning operator at each location $(s,t)$, which shall later allow us to conclude on its stationary marginal distribution, see Section~\ref{subsection: Stochastic Properties of CINAR Random Fields} for details. 

% {\color{red}
% Moreover, note that here, we consider a unilateral model structure, i.e., we propose a natural ordering to the sites. This is reasonable, e.g., when modeling real spatial data which has a slope of the ground or wind coming from a specific direction as an underlying mechanism. Examples include the number of plants affected by fungal disease or pests which spread predominantly with prevailing winds, seeds that are washed downhill after a heavy rain, or the number of clogged drains in a stormwater drainage systems in urban areas. Another prominent application is given by migrational data. People might flee in one direction, e.g, because of war, famine or natural disasters. The random field could then be interpreted as a snapshot that explains the distribution of the population across a grid. Models where each observation is affected by all adjacent values require a multilateral architecture, which is briefly discussed in Section~\ref{subsection: Outlook on Possible Extensions}.
% }

\begin{remark}
\label{remark: independence assumptions}
    Comparing Definition~\ref{def: CINAR} with the corresponding definition of the CINAR time-series model in \citet{weiss08c}, we note slightly different assumptions on the joint distribution of the involved thinnings. The CINAR random-field model is defined by assuming that the thinnings \emph{at ``time''} $(s,t)$ are performed independently of the ``past'' $\mathcal{P}\!_{s,t}$, whereas the CINAR time-series model would correspond to assuming the thinnings \emph{being applied to} $X_{s,t}$ (at ``later times'') to be independent of the ``past'' $\mathcal{P}\!_{s,t}$. In the latter setup, different assumptions on the joint distribution of the thinnings would be possible, which would then lead to different model types in analogy to \citet{weiss08c}. The former setup (the one chosen by us in Definition~\ref{def: CINAR}), by contrast, is analogous to the ``independent thinnings'' approach in \citet{weiss08c}, and it has also been considered by \citet{sil_wei_26} in the context of ordinary INAR random fields. Here, we adapt the approach of \citet{sil_wei_26} to ensure the tractability of the model properties as well as the comparability between INAR and CINAR models for count random fields. We refer the reader to \citet[Remark~2]{sil_wei_26} for a more detailed discussion.
\end{remark}

\subsection{Stochastic Properties of CINAR Random Fields}
\label{subsection: Stochastic Properties of CINAR Random Fields}
Let us start our investigations with a general result on the ergodicity of CINAR random fields.

\begin{proposition}
\label{prop: ergodicity}
    A stationary CINAR random field according to Definition~\ref{def: CINAR} is ergodic.
\end{proposition}

The proof of Proposition~\ref{prop: ergodicity}, which is presented in Appendix~\ref{Proof of Proposition prop: ergodicity}, employs standard techniques in analogy to, e.g., the corresponding proof of \citet{ghodsi15} for the INAR$(1,1)$ random field.

\smallskip
While the definition in \eqref{eq: CINAR recursion} might look somewhat artificial at first glance, its main advantage becomes clear with the following result on the stationary marginal distribution of $(X_{s,t})$. There, the marginal distribution is expressed in terms of the probability generating function (pgf) of~$X_{s,t}$, i.e., of $\pgf_X(u)=\bbe(u^{X_{s,t}})$, which uniquely encodes the probability mass function (PMF) of~$X_{s,t}$.

\begin{proposition}
    \label{prop: CINAR marginal properties}
    Let $(X_{s,t})$ be a stationary CINAR random field according to Definition~\ref{def: CINAR}. Then, its marginal distribution satisfies
    \begin{equation}
        \label{eq: CINAR pgf}
        \pgf_X(u) = \pgf_X(1-\alpha+\alpha u) \cdot \pgf_\varepsilon(u),
    \end{equation}
    where mean and variance are given by
    \begin{equation}
        \label{eq: CINAR mean and var}
        \mu_X = \frac{\mu_\varepsilon}{1-\alpha} \quad \text{and} \quad \sigma^2_X = \frac{\alpha\, \mu_\varepsilon + \sigma^2_\varepsilon}{1-\alpha^2} = \mu_X\,\frac{\alpha+\sigma^2_\varepsilon/\mu_\varepsilon}{1+\alpha}, \quad\text{respectively}.
    \end{equation}
\end{proposition}
The proof of Proposition~\ref{prop: CINAR marginal properties} is presented in Appendix~\ref{Proof of Proposition prop: CINAR marginal properties}. Note that \eqref{eq: CINAR pgf} corresponds to the pgf recursion of the classical INAR(1) model, i.e., the CINAR$(p_1,p_2)$ model can be equipped with exactly those marginal distributions that can also be attained by the classical INAR$(1)$ model \citep[Section~2.1.3]{weiss18}. In particular, in analogy to the INMA random field discussed in \citet{sil_wei_26b}, compound Poisson (CP)-distributed innovations~$\varepsilon_{s,t}$ lead to CP-distributed observations~$X_{s,t}$. Furthermore, we can choose each member of the family of discrete self-decomposable (DSD) distributions as a marginal distribution of~$X_{s,t}$, where the required innovations' distribution is then given by $\pgf_\varepsilon(u) = \pgf_X(u)\, /\, \pgf_X(1-\alpha+\alpha u)$. The DSD family includes, among others, the Poi- and negative-binomial (NB) distribution (see \citet{zhu03,weiss08b} as well as Example~\ref{example: Poi NB marginal} below for further details), which are also later considered in our simulation study presented in Section~\ref{section: Performance of Parameter Estimation}. Equation \eqref{eq: CINAR mean and var}, in turn, shows that we have over-/equi-/underdispersed observations if and only if (iff) the innovations are over-/equi-/underdispersed.

\smallskip
Next, we investigate the spatial dependence structure of the stationary CINAR random field $(X_{s,t})$. For $k,l\in\bbz$, its spatial ACF $\rho$ is defined by
\[
    \rho(k,l) = \corr(X_{s,t}, X_{s+k,t+l}) =  \frac{\cov(X_{s,t}, X_{s+k,t+l})}{\var(X_{s,t})}.
\]
Note that $\rho(k,l)=\rho(-k,-l)$ by definition.
As shown in the following proposition, the CINAR's ACF satisfies a set of Yule--Walker (YW) equations.

\begin{figure}[t]
\centering\small
(a)~\includegraphics[viewport=65 550 310 770, clip=, scale=0.55]{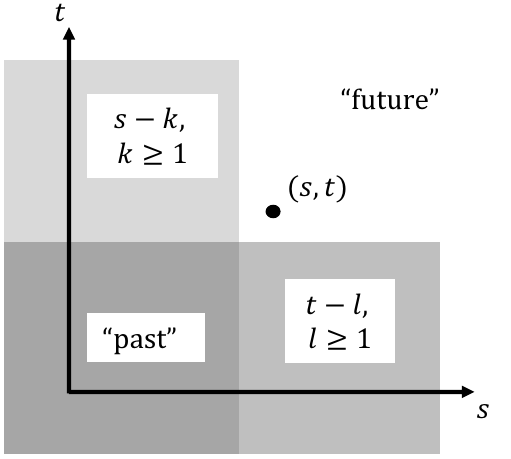}
\qquad
(b)\hspace{-3ex}\includegraphics[viewport=80 580 290 765, clip=, scale=0.65]{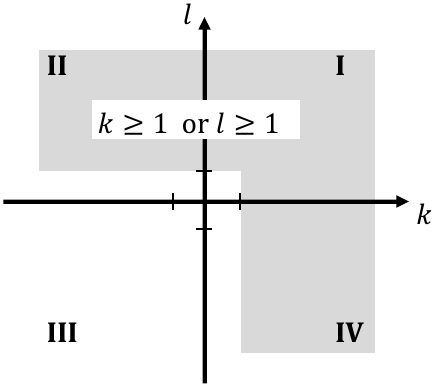}
\qquad
(c)\hspace{-3ex}\includegraphics[viewport=65 580 275 765, clip=, scale=0.65]{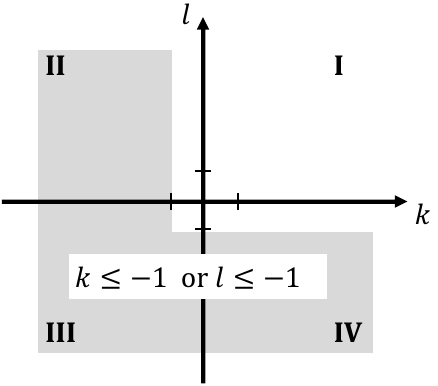}
\caption{(a) Process' ``past'' corresponding to ``time'' $(s,t)$. Spatial lags $(k,l)$ and Quadrants I--IV, where (b) equation \eqref{eq: CINAR sacf recursion 1} or (c) equation \eqref{eq: CINAR sacf recursion 2} holds. Graphs adapted from \citet[Fig.\ 1]{sil_wei_26}}.
\label{figAreas}
\end{figure}

\begin{proposition}
    \label{prop: CINAR spatial properties}
    Let $(X_{s,t})$ be a stationary CINAR random field according to Definition~\ref{def: CINAR}. Then, its ACF satisfies the recursions
    \begin{align}
    \rho(k,l) &= \alpha \sum_{(i,j)\in\mathcal{S}} \phi_{ij}\, \rho(k-i,l-j) \qquad \text{if } k\geq 1 \text{ or } l\geq 1, \label{eq: CINAR sacf recursion 1} \\
    \rho(k,l) &= \alpha \sum_{(i,j)\in\mathcal{S}} \phi_{ij}\, \rho(k+i,l+j) \qquad \text{if } k\leq -1 \text{ or } l\leq -1. \label{eq: CINAR sacf recursion 2}
\end{align}
\end{proposition}
The proof of Proposition~\ref{prop: CINAR spatial properties} is presented in Appendix~\ref{Proof of Proposition prop: CINAR spatial properties}. Note that Equation \eqref{eq: CINAR sacf recursion 1} holds in the region illustrated by Figure~\ref{figAreas}\,(b), and Equation \eqref{eq: CINAR sacf recursion 2} in the one of Figure~\ref{figAreas}\,(c). Also note that the YW~equations of Proposition~\ref{prop: CINAR spatial properties} agree with those of the ordinary INAR$(p_1,p_2)$ random field after the following reparametrization: if one defines $\theta_{ij} = \alpha\cdot\phi_{ij}$ with $\sum_{(i,j)\in\mathcal{S}} \theta_{ij} =\alpha$, then the~$\theta_{ij}$ are the counterparts to the AR coefficients~$\alpha_{ij}$ in \eqref{eq: generalized model}. In particular, replacing~$\theta_{ij}$ in Proposition~\ref{prop: CINAR spatial properties} by~$\alpha_{ij}$, we obtain Proposition~5 of \citet{sil_wei_26}. 
Hence, altogether, the CINAR$(p_1,p_2)$ approach combines two highly attractive features, namely the classical AR-type autocorrelation structure with a broad variety of possible marginal distributions. So recalling George Box' famous words ``All models are wrong, but some are useful'', we clearly judge the CINAR models as being useful for practice despite their somewhat technical definition.

\smallskip
A further advantage of the CINAR compared to the ordinary INAR approach becomes clear when studying its regression properties. 

\begin{proposition}
    \label{prop: CINAR cond distr}
    Let $(X_{s,t})$ be a stationary CINAR random field according to Definition~\ref{def: CINAR}. Then, the conditional distribution is given by 
    \begin{equation}
        \label{eq: CINAR cond dist}
        \bbp(X_{s,t}=x\, |\, \mathcal{P}\!_{s,t}) 
        = \sum^x_{y=0} \bbp(\varepsilon_{s,t}=y) \cdot \sum_{(i,j)\in\mathcal{S}} \phi_{ij} \cdot \binom{X_{s-i,t-j}}{x-y} \alpha^{x-y} (1-\alpha)^{X_{s-i,t-j}-x+y},
    \end{equation}
    where the binomial probabilities $\binom{X_{s-i,t-j}}{x-y} \alpha^{x-y} (1-\alpha)^{X_{s-i,t-j}-x+y}$ are equal to zero if $x-y>X_{s-i,t-j}$. 
    In particular, the conditional mean and variance are given by
    \begin{equation}
        \label{eq: CINAR cond exp}
        \bbe(X_{s,t}\, |\, \mathcal{P}\!_{s,t}) = \mu_\varepsilon + \alpha \sum_{(i,j)\in\mathcal{S}} \phi_{ij} X_{s-i,t-j}
    \end{equation}
    and
    \begin{equation}
        \label{eq: CINAR cond var}
        \var(X_{s,t}\, |\, \mathcal{P}\!_{s,t}) = \sigma^2_\varepsilon + \alpha(1-\alpha) \sum_{(i,j)\in\mathcal{S}} \phi_{ij} X_{s-i,t-j} + \alpha^2 \sum_{(i,j)\in\mathcal{S}} \phi_{ij} X_{s-i,t-j}^2 - \alpha^2 \sum_{(i,j),(k,l)\in\mathcal{S}} \phi_{ij} \phi_{kl} X_{s-i,t-j} X_{s-k,t-l},
    \end{equation}
    respectively.
\end{proposition}
The proof of Proposition~\ref{prop: CINAR cond distr} is presented in Appendix~\ref{Proof of Proposition prop: CINAR cond distr}. Note that Proposition~\ref{prop: CINAR cond distr} is of utmost importance for statistical inference. First, it can be utilized for parameter estimation, namely the conditional mean \eqref{eq: CINAR cond exp} for conditional least squares (CLS) estimation and the conditional distribution \eqref{eq: CINAR cond dist} for conditional maximum likelihood (CML) estimation, see Section~\ref{section: Parameter Estimation} for details. Secondly, Proposition~\ref{prop: CINAR cond distr} is useful for model diagnostics, as it allows to compute the standardized Pearson residuals and the probability integral transform (PIT) histogram, see \citet[Section~2.4]{weiss18} and \citet{jung22} for details. These diagnostic tools shall also later be used in our data application presented in Section~\ref{Data Example: Yields from an Agricultural Experiment}.

\smallskip
As mentioned before, Proposition~\ref{prop: CINAR cond distr} shows a further major advantage of the CINAR compared to the INAR approach: the conditional probabilities \eqref{eq: CINAR cond dist} are much simpler to compute than those of classical INAR random fields, as we are only concerned with one convolution as caused by the single selected binomial thinning at location $(s,t)$. By contrast, the INAR$(1,1)$ random field requires a three-fold convolution \citep[see][Proposition 2.1]{ghodsi15}, and the INAR$(p_1,p_2)$ random field more generally an $|\mathcal{S}|$-fold convolution \citep[see][Proposition 7]{sil_wei_26}. Also the multilateral INAR$(1,1)$ model according to \citet{karlis24} is computationally much more demanding, as it requires an eight-fold convolution according to their Equation (9). Note that a brief discussion on multilateral CINAR models is provided in Section~\ref{subsection: Outlook on Possible Extensions} below. But first, let us take a closer look at the case of the first-order CINAR model.

\subsection{Special Case: The CINAR$(1,1)$ Model}
\label{subsection: Special Case: The CINAR(1,1) Model}
Up to now, the vast majority of articles on count random fields focused on the first-order INAR model, i.e., on model \eqref{eq: model} and its modifications, recall the literature review in Section~\ref{section: Introduction}. In fact, this special case often turned out to be sufficient for modeling real-world count grid data, and it was also possible to derive closed-form expressions for its spatial ACF. However, as already mentioned, even for these first-order INAR models like \eqref{eq: model}, it was not possible to characterize the stationary marginal distribution beyond expressions for mean and variance. Therefore, the present section presents a brief overview on essential stochastic properties of the CINAR$(1,1)$ model, defined by the recursion
\begin{equation}
    \label{eq: cinar11 model}
    X_{s,t} = D_{s,t;0,1}\, (\alpha \circ X_{s,t-1}) + D_{s,t;1,0}\, (\alpha \circ X_{s-1,t}) + D_{s,t;1,1}\, (\alpha \circ X_{s-1,t-1})\ +\ \varepsilon_{s,t},
\end{equation}
as a natural competitor to the existing INAR$(1,1)$ models. The corresponding marginal distribution is characterized in Proposition~\ref{prop: CINAR marginal properties}, which applies to CINAR models of any order in the same way. In particular, as for any model order $(p_1,p_2)$, it is possible to achieve Poi- or NB-distributed outcomes~$X_{s,t}$.

\begin{example}
\label{example: Poi NB marginal}
If the innovations $(\varepsilon_{s,t})$ of a CINAR random field (of any order) are i.i.d.\ Poi-distributed according to $\poi(\mu_\varepsilon)$, then the observations $X_{s,t}$ are Poi-distributed as well (and thus equidispersed, i.e., $\sigma_X^2=\mu_X$), namely $X_{s,t}\sim\poi(\mu_X)$ with $\mu_X = \mu_\varepsilon/(1-\alpha)$, see \citet[Example~3.3]{weiss08b}. We refer to this special case as the Poi-CINAR model.

\smallskip
If, by contrast, we want to achieve the (overdispersed) NB-distribution for $X_{s,t}$, say, $X_{s,t}\sim\nb(\nu,\pi)$ with $\nu>0$ and $\pi\in (0,1)$ such that $\mu_X=\nu(1-\pi)/\pi$ and $\sigma_X^2/\mu_X = 1/\pi\ >1$, then the innovations must have the following PMF \citep[Example~3.4]{weiss08b}:
$$
P(\varepsilon_{s,t}=k)\ =\ \left\{
\begin{array}{ll}
\big(1-(1-\pi)(1-\alpha)\big)^\nu & \text{if } k=0,\\[2mm]
\sum\limits_{j=1}^k\ \binom{\nu}{j}\, \big(1-(1-\pi)(1-\alpha)\big)^{\nu-j}\, \big((1-\pi)(1-\alpha)\big)^j\ \cdot\ \binom{k-1}{k-j}\, (1-\pi)^{k-j}\, \pi^j & \text{if } k>0,
\end{array}
\right.
$$
where the (generalized) binomial coefficient $\binom{\nu}{j}$ is defined by $\Gamma(\nu+1)\,/\,\big(\Gamma(\nu-j+1)\cdot j!\big)$. The mean and variance of~$\varepsilon_{s,t}$ satisfy the equations given in Proposition~\ref{prop: CINAR marginal properties}, and we refer to the model as the NB-CINAR one. 
\end{example}
Propositions~\ref{prop: CINAR spatial properties} and~\ref{prop: CINAR cond distr} simplify for the CINAR$(1,1)$ model by using that $\mathcal{S} = \{(0,1),\ (1,0),\ (1,1)\}$. Since the YW~equations of Proposition~\ref{prop: CINAR spatial properties} agree with those of the stationary INAR$(p_1,p_2)$ random field \eqref{eq: generalized model} after replacing~$\theta_{ij}=\alpha\,\phi_{ij}$ by~$\alpha_{ij}$ (see \citealp{sil_wei_26}, Proposition~5), we also obtain the same solution in the special case $p_1=p_2=1$ as for the ordinary INAR$(1,1)$ random field, i.e., the same closed-form expression for the ACF. This expression is provided by Proposition~3.2 of \citet{ghodsi12} as well as Proposition~2 and Corollary~1 of \citet{sil_wei_26}, and it remains valid after replacing their~$\alpha_{ij}$ by our~$\theta_{ij}$, which leads to the following corollary.

\begin{corollary}
\label{cor: cinar11 acf}
    Let $(X_{s,t})$ be a stationary CINAR$(1,1)$ random field according to Definition~\ref{def: CINAR} and Equation \eqref{eq: cinar11 model}, let $k,l\geq 0$. Then, the ACF over Quadrants~II and~IV (recall Figure~\ref{figAreas}) is given by
    \begin{equation*}
        \rho(-k,l) = \rho(k,-l) = \lambda^k \eta^{l}
        \quad\text{with } \lambda,\eta\in(0,1),
    \end{equation*}
    where
    \begin{equation*}
        \begin{split}
            \lambda &= \frac{1+\alpha^2(\phi_{10}^2-\phi_{01}^2-\phi_{11}^2) - \sqrt{\big(1+\alpha^2(\phi_{10}^2-\phi_{01}^2-\phi_{11}^2)\big)^2 - 4\alpha\, (\phi_{10}+\alpha\, \phi_{01} \phi_{11})}}{2\alpha (\phi_{10}+\alpha\, \phi_{01} \phi_{11})} 
            % \\
            % &
            \quad
            \text{and} \quad \eta=\frac{\alpha\, (\phi_{01}+\phi_{11}\lambda)}{1-\alpha\, \phi_{10} \lambda}.
        \end{split}
    \end{equation*}
    Furthermore, the ACF over Quadrants~I and~III, $\rho(-k,-l) = \rho(k,l)$, satisfies $\rho(k,0) = \lambda^k$, $\rho(0,l) = \eta^{l}$, and
    $$
    \rho(k,l) = \alpha\,\phi_{01}\, \rho(k,l-1) + \alpha\,\phi_{10}\, \rho(k-1,l) + \alpha\,\phi_{11}\, \rho(k-1,l-1)
    \quad\text{for } k,l\geq 1.
    $$
\end{corollary}
To sum up, the CINAR$(1,1)$ random field exhibits the same autocorrelation structure as existing INAR$(1,1)$ random fields, but it can be easily equipped with common marginal distributions, and also its conditional probabilities are simpler to compute. Thus, the CINAR$(1,1)$ model is an appealing competitor of existing INAR$(1,1)$ models for count random fields.

\subsection{Outlook on Possible Extensions}
\label{subsection: Outlook on Possible Extensions}
Let us conclude Section~\ref{section: The Combined INAR Random Field} by briefly discussing possible future extensions of our novel CINAR$(p_1,p_2)$ model for count random fields. First, in analogy to the multilateral INAR$(1,1,1,1)$ model of \citet{karlis24}, one could define a multilateral CINAR recursion of order $(p_1,p_2,q_1,q_2)\in\bbn^4$ as
    \begin{equation}
        \label{eq: multilateral CINAR recursion}
        X_{s,t}
        = \sum_{(i,j)\in\mathcal{S}^*} D_{s,t;i,j} \cdot (\alpha \circ X_{s-i,t-j}) + \varepsilon_{s,t}
    \end{equation}
similar to Definition~\ref{def: CINAR}, where $\mathcal{S}^*:=\big\{(i,j) \,|\, -q_1 \leq i \leq p_1,\, -q_2 \leq j \leq p_2,\, (i,j) \neq (0,0)\big\}$, and where the ``past'' $\mathcal{P}\!_{s,t}$ is replaced by $\mathcal{A}_{s,t} = \{X_{u,v}: (u,v)\not=(s,t)\}$, i.e.,  all random variables other than~$X_{s,t}$. Then, provided that the existence of a stationary solution can be established, the marginal properties according to Proposition~\ref{prop: CINAR marginal properties} could be shown to hold for the multilateral CINAR$(p_1,p_2,q_1,q_2)$ recursion as well by simply replacing the summation across~$\mathcal{S}$ in the proof in Appendix~\ref{Proof of Proposition prop: ergodicity} by a summation across~$\mathcal{S}^*$. In a similar way, the conditional probability \eqref{eq: CINAR cond dist} in Proposition~\ref{prop: CINAR cond distr} would change to 
\begin{equation}
    \label{eq: multilateral CINAR cond dist}
    \bbp(X_{s,t}=x\, |\, \mathcal{A}_{s,t}) 
    = \sum^x_{y=0} \bbp(\varepsilon_{s,t}=y) \cdot \sum_{(i,j)\in\mathcal{S}^*} \phi_{ij} \cdot \binom{X_{s-i,t-j}}{x-y} \alpha^{x-y} (1-\alpha)^{X_{s-i,t-j}-x+y},
\end{equation}
which again involves only a single convolution. Formula \eqref{eq: multilateral CINAR cond dist} is computationally much cheaper than Equation (9) in \citet{karlis24}, because their conditional probability involves an eight-fold convolution instead. While a comprehensive analysis of its stochastic properties is certainly still pending (in particular, parameter estimation is expected to be demanding in view of \citealp{whittle54}), the above illustrations already indicate that a multilateral CINAR model defined via \eqref{eq: multilateral CINAR recursion} could become an appealing competitor of the multilateral INAR model developed by \citet{karlis24}. Thus, a detailed investigation of a multilateral CINAR model is recommended for future research.

\bigskip
Second, one could extend the CINAR random field to negative dependencies by combining it with the Tobit approach developed by \citet{wei_zhu_kim_26} in the time-series case. More precisely, with the same assumptions as in Definition~\ref{def: CINAR}, the Tobit CINAR$(p_1,p_2)$ model for count random fields would be defined by
\begin{equation}
    \label{eq: tobit CINAR recursion}
    X_{u,v} = \max\{0,\ Y_{u,v}\}
    \quad\text{with}\quad
    Y_{u,v} = \sum_{(i,j)\in\mathcal{S}} D_{u,v;i,j} \cdot s(i,j) \cdot (\alpha \circ X_{u-i,v-j}) + \varepsilon_{u,v},
\end{equation}
where $s(i,j) \in\{\pm 1\}$ denotes the specified sign of the $(i,j)$\textsuperscript{th} autoregressive term. Here, the censoring at zero avoids that $X_{u,v}$ becomes negative, whereas~$Y_{u,v}$ is allowed to do so. For example, a Tobit CINAR$(1,1)$ model might be defined by
$$
    X_{u,v} = \max\{0,\ Y_{u,v}\}
    \quad\text{with}\quad
    Y_{u,v} = D_{u,v;0,1} \cdot (\alpha \circ X_{u,v-1}) + D_{u,v;1,0} \cdot (\alpha \circ X_{u-1,v}) - D_{u,v;1,1} \cdot (\alpha \circ X_{u-1,v-1}) + \varepsilon_{u,v}
$$
if one chooses $s(0,1)=s(1,0)=+1$ and $s(1,1)=-1$. For the general Tobit CINAR$(p_1,p_2)$ model \eqref{eq: tobit CINAR recursion}, the conditional distribution $\bbp(X_{u,v}=x\, |\,\mathcal{P}\!_{u,v})$ is computed in two steps. Abbreviate the PMF of the $\bin(n,\pi)$-distribution by $\mathfrak{p}_{n,\pi}(z)$ for $z\in\bbz$, which equals zero if $z\not\in\{0,\ldots,n\}$. Then,
\begin{equation}
    \label{eq: tobit CINAR cond dist y}
    \bbp(Y_{u,v}=x\, |\,\mathcal{P}\!_{u,v}) = \sum_{(i,j)\in\mathcal{S}} \phi_{ij} \cdot \sum^\infty_{y=0} \bbp(\varepsilon_{u,v}=y) \cdot \mathfrak{p}_{X_{u-i,v-j},\alpha}\brackets{s(i,j) \cdot (x-y)}
    \quad\text{for } x\in\bbz,
\end{equation}
where for each $(i,j)\in\mathcal{S}$, the inner summation across~$y$ has only finitely many non-zero summands due to the finite support of~$\mathfrak{p}_{X_{u-i,v-j},\alpha}(z)$. For the same reason, there always exists a finite lower bound $L_{u,v}$ for the support of $\bbp(Y_{u,v}=x\, |\,\mathcal{P}\!_{u,v})$, namely $L_{u,v} := \sum_{(i,j)\in\mathcal{S}} \min\big\{0, s(i,j)\big\}\cdot X_{u-i,v-j}$. 
Then,
\begin{equation}
    \label{eq: tobit CINAR cond dist x}
    \bbp(X_{u,v}=x\, |\,\mathcal{P}\!_{u,v}) = \left\{\begin{array}{ll}
    \bbp(Y_{u,v}=x\, |\,\mathcal{P}\!_{u,v}) & \text{if } x>0, \\[1ex]
    \sum_{z=L_{u,v}}^0 \bbp(Y_{u,v}=z\, |\,\mathcal{P}\!_{u,v}) & \text{if } x=0.
    \end{array}\right.
\end{equation}
Formulas \eqref{eq: tobit CINAR cond dist y}--\eqref{eq: tobit CINAR cond dist x} are proven in Appendix~\ref{Proof of Equations eq: tobit CINAR cond dist y eq: tobit CINAR cond dist x}, and they would be useful for likelihood inference (among others). In analogy to the Tobit INAR time-series model of \citet{wei_zhu_kim_26}, the Tobit CINAR$(p_1,p_2)$ model \eqref{eq: tobit CINAR recursion} does not behave \emph{exactly} like a linear model (due to the censoring at zero), but it is reasonable to assume that, despite the presence of some negative AR terms, its dependence structure is often at least \emph{approximately} linear. This conjecture as well as further stochastic properties of the Tobit CINAR$(p_1,p_2)$ model \eqref{eq: tobit CINAR recursion} should be analyzed in future research. Finally, opportunities for incorporating
covariate information into the CINAR framework (or the aforementioned extensions) constitute a relevant direction for future research.

\section{Parameter Estimation}
\label{section: Parameter Estimation}
In this section, we deal with methods for parameter estimation, where all estimators are based on the rectangular segment $X_{1,1}, \dots, X_{n_1,n_2}$ from $(X_{s,t})$ with $n_1,n_2 \in \bbn$. 
For the stationary CINAR random field model, parameter estimation can be done by inserting the empirical instead of the theoretical autocorrelations into the respective YW~equations (recall Proposition~\ref{prop: CINAR spatial properties}) and solving them, which leads to the YW~estimates. Moreover, the conditional expectation and distribution according to Proposition~\ref{prop: CINAR cond distr} can be used to compute the CLS and CML estimates. All three estimation approaches have also been considered for the classical INAR$(1,1)$ random field model, see \citet{ghodsi12}, \citet{ghodsi15}, and \citet{sassi23}. For the sake of uniqueness and comparability to existing INAR results, we consider the reparametrization $\theta_{ij}=\alpha \cdot \phi_{ij}$, i.e., we estimate the parameters $\bftheta=(\theta_{01}, \dots, \theta_{p_1,p_2})^\top$ (note the lexicographic ordering of the parameters~$\theta_{ij}$ within $\bftheta$), where $\sum_{(i,j)\in\mathcal{S}} \theta_{ij} = \alpha$ and $\phi_{ij} = \theta_{ij}/\alpha$. In addition, we estimate the relevant parameters of the innovations process, such as their mean $\mu_\varepsilon$ and possibly the variance $\sigma_\varepsilon^2$. Note that if the innovations follow a Poi-distribution, then the innovations are already fully characterized by solely~$\mu_\varepsilon$.

%%%%%
\subsection{Yule--Walker Estimation}
\label{subsection: Yule--Walker Estimation}

It is well-known that YW~estimators are based on the ACF of the respective model. We have shown that for $p_1=p_2=1$, the ACF of the CINAR random field coincides with the ACF of the ordinary INAR$(1,1)$ random field if the coefficients $\theta_{ij}= \alpha \phi_{ij}$ are replaced by $\alpha_{ij}$, recall Section~\ref{subsection: Special Case: The CINAR(1,1) Model}. Consequently, we expect strong similarities to the YW~estimator for the INAR$(1,1)$ random field, which has been derived by \citet{ghodsi12}. Nevertheless, we present the derivation in order to account for the higher model orders. It should be noted, however, that the YW~estimator derived here extends directly to higher-order INAR random fields \eqref{eq: generalized model} according to \citet{sil_wei_26} (if the coefficients are replaced accordingly). 
From \eqref{eq: CINAR sacf recursion 1}, we know that
\begin{equation}
\label{eq: YW matrix equation}
    \boldsymbol{\rho} = \mathbf{P} \cdot \bftheta ,
\end{equation}
where $\bftheta=(\theta_{ij})_{(i,j)\in\mathcal{S}}$, $\boldsymbol{\rho}=(\rho(k,l))_{(k,l)\in\mathcal{S}}$ and $\mathbf{P}=(\rho(k-i,l-j))_{(k,l),(i,j)\in\mathcal{S}}$. Since $\rho(k,l)=\rho(-k,-l)$ for all $k,l\in\bbz$, we replace $\rho(k-i,l-j)$ with $\rho(-k+i,-l+j)$ if $k-i<0$, and $\rho(0,l-j)$ with $\rho(0,-l+j)$ if $k-i=0$ and $l-j<0$.
This way, only autocorrelations $\rho(k,l)$ with $(k,l)\in\mathcal{S}$ or $k\in\{1, \dots, p_1\}$ and $l\in\{-1,\dots, -p_2\}$ appear in \eqref{eq: YW matrix equation}. 
In case of a first-order model ($p_1=p_2=1$), this yields $\boldsymbol{\rho} = (\rho(0,1), \rho(1,0), \rho(1,1))^\top$ and
\[
    \mathbf{P} = \begin{pmatrix}
        1 & \rho(-1,1) & \rho(-1,0) \\
        \rho(1,-1) & 1 & \rho(0,-1) \\
        \rho(1,0) & \rho(0,1) & 1
    \end{pmatrix}
    = \begin{pmatrix}
        1 & \rho(1,-1) & \rho(1,0) \\
        \rho(1,-1) & 1 & \rho(0,1) \\
        \rho(1,0) & \rho(0,1) & 1
    \end{pmatrix}.
\]
Note that the difference to \citet{ghodsi12} comes from a different ordering of the coefficients (we uniquely use a lexicographic ordering). After replacing all autocorrelations $\rho(k,l)$ in $\boldsymbol{\rho}$ and $\mathbf{P}$ from \eqref{eq: YW matrix equation} by their sample analogs $\hat{\rho}(k,l) := \hat{\gamma}(k,l)/\hat{\gamma}(0,0)$, we denote this vector and matrix by $\hat{\boldsymbol{\rho}}$ and $\hat{\mathbf{P}}$, respectively, where the sample ACvF $\hat{\gamma}(k,l)$ is defined by
\[
    \hat{\gamma}(k,l) := \frac{1}{n_1 n_2} \sum^{n_1}_{s=1+k}\sum^{n_2-l}_{t=1+l} (X_{s,t} - \overline{X})(X_{s-k,t-l}-\overline{X}) \quad \text{with} \quad \overline{X} = \frac{1}{n_1 n_2} \sum^{n_1}_{s=1} \sum^{n_2}_{t=1} X_{s,t}.
\] 
Now, the YW~estimator $\hat{\bftheta}_{\text{YW}} = (\hat{\theta}_{\text{YW}; ij})_{(i,j)\in\mathcal{S}}$ of $\bftheta$ is obtained by solving
\begin{equation}
\label{eq: YW theta}
    \hat{\bftheta}_{\text{YW}} = \hat{\mathbf{P}}^{-1} \hat{\boldsymbol{\rho}}.
\end{equation}
Therefore, $\hat{\alpha}_{\text{YW}} = \sum_{(i,j)\in\mathcal{S}} \hat{\theta}_{\text{YW}; ij}$ and $\hat{\phi}_{\text{YW}; ij} = \hat{\theta}_{\text{YW}; ij} / \hat{\alpha}_{\text{YW}}$. Moreover, using \eqref{eq: CINAR mean and var}, $\mu_\varepsilon$ and $\sigma^2_\varepsilon$ are estimated by 
\begin{equation}
\label{eq: YW inno}
    \hat{\mu}_{\text{YW};\, \varepsilon} = \overline{X} \cdot (1- \hat{\alpha}_{\text{YW}}) \qquad \text{and} \qquad \hat{\sigma}_{\text{YW};\, \varepsilon}^2 = \hat{\sigma}_X^2 \cdot (1-\hat{\alpha}_{\text{YW}}) - \hat{\alpha}_{\text{YW}} \cdot \hat{\mu}_{\text{YW};\, \varepsilon},
\end{equation}
where $\hat{\sigma}_X^2:=\hat{\gamma}(0,0)$. Note that while the definition of the YW~estimator is intuitive, it may lead to inadmissible estimates in practice, i.e., estimates that violate the parameter constraints according to Definition~\ref{def: CINAR}.

%%%%%
\subsection{Conditional Least Squares Estimation}
\label{subsection: Conditional Least Squares Estimation}
The CLS estimator of $\bfvartheta=(\theta_{01}, \dots, \theta_{p_1,p_2}, \mu_\varepsilon)^\top$, the parameters being involved in the conditional mean \eqref{eq: CINAR cond exp}, is given by the statistic $\hat{\bfvartheta}_{\text{CLS}} := \arg \min_{\bfvartheta} \ Q_{n_1,n_2}(\bfvartheta)$ minimizing the sum of squared deviations
\begin{align*}
    Q_{n_1,n_2}(\bfvartheta) := \sum^{n_1}_{s=p_1+1} \sum^{n_2}_{t=p_2+1} (X_{s,t}-\bbe(X_{s,t}|\mathcal{P}\!_{s,t}))^2 = \sum^{n_1}_{s=p_1+1} \sum^{n_2}_{t=p_2+1} \brackets{X_{s,t} - \mu_\varepsilon - \sum_{(i,j)\in\mathcal{S}}\theta_{ij} X_{s-i,t-j}}^2. 
\end{align*}
Besides a direct numerical (constrained) minimization of~$Q_{n_1,n_2}(\bfvartheta)$, also closed-form expressions for the CLS estimator can be derived. The partial derivatives of $Q_{n_1,n_2}(\bfvartheta)$ are given by
\begin{align}
    \frac{\partial}{\partial \mu_\varepsilon} Q_{n_1,n_2}(\bfvartheta) &= -2 \sum^{n_1}_{s=p_1+1} \sum^{n_2}_{t=p_2+1} \brackets{X_{s,t} - \mu_\varepsilon - \sum_{(i,j)\in\mathcal{S}}\theta_{ij} X_{s-i,t-j}} \label{eq: CLS part der 1}, \\
    \frac{\partial}{\partial \theta_{k,l}} Q_{n_1,n_2}(\bfvartheta) &= -2 \sum^{n_1}_{s=p_1+1} \sum^{n_2}_{t=p_2+1} \brackets{X_{s-k,t-l} \cdot \brackets{X_{s,t} - \mu_\varepsilon - \sum_{(i,j)\in\mathcal{S}}\theta_{ij} X_{s-i,t-j}}} \label{eq: CLS part der 2}\quad\text{for } (k,l)\in\mathcal{S}.
\end{align}
For notational convenience, we write $n := (n_1-p_1) (n_2-p_2)$. Equating \eqref{eq: CLS part der 1} to zero, we obtain
\begin{equation}
    \label{eq: CLS mu}
    \mu_\varepsilon = \frac{1}{n} \sum^{n_1}_{s=p_1+1} \sum^{n_2}_{t=p_2+1} \brackets{X_{s,t} - \sum_{(i,j)\in\mathcal{S}}\theta_{ij} X_{s-i,t-j}}.
\end{equation}
Then, equating \eqref{eq: CLS part der 2} to zero and substituting $\mu_\varepsilon$, we get 
\begin{equation}
    \label{eq: CLS thetakl} 
    \sum_{(i,j)\in\mathcal{S}} \theta_{ij}\, \tilde{\gamma}((i,j),(k,l)) = \tilde{\gamma}((k,l), (0,0)), \quad (k,l)\in\mathcal{S},
\end{equation}
where
\[
    \tilde{\gamma}((i,j),(k,l)) := \brackets{\frac{1}{n} \sum^{n_1}_{s=p_1+1} \sum^{n_2}_{t=p_2+1} X_{s-i,t-j} \cdot X_{s-k,t-l}} - \brackets{\frac{1}{n} \sum^{n_1}_{s=p_1+1} \sum^{n_2}_{t=p_2+1} X_{s-i,t-j}} \cdot \brackets{\frac{1}{n} \sum^{n_1}_{s=p_1+1} \sum^{n_2}_{t=p_2+1} X_{s-k,t-l}}.
\]
Combining \eqref{eq: CLS mu} and \eqref{eq: CLS thetakl} is equivalent to solving 
\begin{equation}
\label{eq: eq system CLS}
    \mathbf{A} \cdot \boldsymbol{\vartheta} = \mathbf{b},
\end{equation}
where $\mathbf{b}$ is the $(|\mathcal{S}|+1)=(p_1+1)(p_2+1)$-dimensional vector defined by
\[
    \mathbf{b} = \sum^{n_1}_{s=p_1+1} \sum^{n_2}_{t=p_2+1} ((X_{s,t}\cdot X_{s-i,t-k})_{(i,j)\in\mathcal{S}},\ X_{s,t})^\top,
\]
and $\mathbf{A}$ is the symmetric $((p_1+1)(p_2+1))\times((p_1+1)(p_2+1))$-dimensional matrix
\[
    \mathbf{A} = \sum^{n_1}_{s=p_1+1} \sum^{n_2}_{t=p_2+1} 
    \left(\begin{array}{cc}
        (X_{s-i,t-j} \cdot X_{s-k,t-l})_{(i,j),(k,l)\in\mathcal{S}} & (X_{s-i,t-j})_{(i,j)\in\mathcal{S}}  \\
        (X_{s-k,t-l})_{(k,l)\in\mathcal{S}} & 1
    \end{array}\right),
\]
which contains the matrix $(X_{s-i,t-j} \cdot X_{s-k,t-l})_{(i,j),(k,l)\in\mathcal{S}}$ in the upper left corner.
Hence, the CLS estimator $\hat{\bfvartheta}_{\text{CLS}} = \hat{\bfvartheta}_{\text{CLS}}(n)$ of $\bfvartheta$ is computed from 
\begin{equation*}
    \hat{\bfvartheta}_{\text{CLS}} = \mathbf{A}^{-1} \mathbf{b}.
\end{equation*}

Note that these statistics truly minimize $Q_{n_1,n_2}(\bfvartheta)$, since all second derivatives are positive except for trivial cases (recall that both $(X_{s,t})$ and $(\varepsilon_{s,t})$ are non-negative counts):
\begin{align*}
    \frac{\partial^2}{\partial \theta_{kl} \partial \mu_\varepsilon} Q_{n_1,n_2}(\bfvartheta) &= + 2 \sum_{s=p_1+1}^{n_1} \sum_{t=p_2+1}^{n_2} X_{s-k,t-l} > 0, \\
    \frac{\partial^2}{\partial \theta_{ij} \partial \theta_{kl}} Q_{n_1, n_2}(\bfvartheta) &= + 2 \sum_{s=p_1+1}^{n_1} \sum_{t=p_2+1}^{n_2} X_{s-k,t-l} X_{s-i,t-j} >0, \\
    \frac{\partial^2}{\partial \mu_\varepsilon^2} Q_{n_1, n_2}(\bfvartheta) &= + 2 n > 0.
\end{align*}
Compare these estimators with those of \citet{sassi23}, who have proposed a CLS estimation procedure for INAR$(1,1)$ random field models: $\hat{\boldsymbol{\vartheta}}_{\text{CLS}}$ matches their estimator if $p_1=p_2=1$ and if $\theta_{ij}$ is substituted by $\alpha_{ij}$.
Let us stress that the conditional expectations of the CINAR$(p_1,p_2)$ and INAR$(p_1,p_2)$ models coincide in general if $\theta_{ij}= \alpha\, \phi_{ij}$ is replaced by $\alpha_{ij}$, see \citet{sil_wei_26}. Therefore, $\hat{\boldsymbol{\vartheta}}_{\text{CLS}}$ transfers directly to INAR$(p_1,p_2)$ models with appropriately adjusted coefficients. Finally, $\hat{\alpha}_{\text{CLS}} = \sum_{(i,j)\in\mathcal{S}} \hat{\theta}_{\text{CLS}; ij}$ and $\hat{\phi}_{\text{CLS}; ij} = \hat{\theta}_{\text{CLS}; ij} / \hat{\alpha}_{\text{CLS}}$ as before.

%%%%%%
\subsection{Conditional Maximum Likelihood Estimation}
\label{subsection: Conditional Maximum Likelihood Estimation}
In order to apply the CML procedure, the CINAR model has to be fully specified, i.e., we also need to specifiy the innovations' distribution. For example, in the case of a Poi-CINAR model with $\varepsilon_{s,t}\sim\text{Poi}(\mu_\varepsilon)$, the full model parametrization is given by $\bfvartheta=(\theta_{01}, \dots, \theta_{p_1,p_2}, \mu_\varepsilon)^\top$, whereas the NB-CINAR model requires an additional dispersion parameter, e.g., $\bfvartheta=(\theta_{01}, \dots, \theta_{p_1,p_2}, \mu_\varepsilon, I_\varepsilon)^\top$ with $I_\varepsilon = \sigma_\varepsilon^2/\mu_\varepsilon >1$, recall Example~\ref{example: Poi NB marginal}. 
In order to express the conditional likelihood function of the CINAR$(p_1,p_2)$ random field, we have to split the available rectangular data segment $\mathcal{X}_{1,1}^{n_1,n_2} := \{X_{1,1},\ldots,X_{n_1,n_2}\}$ into two parts, namely into ``upper-right rectangle'' $\mathcal{X}_{p_1+1,p_2+1}^{n_1,n_2} := \{X_{p_1+1,p_2+1},\ldots,X_{n_1,n_2}\}$ and the remaining ``bottom-left boundary'' $\mathcal{X}_{1,1}^{n_1,n_2}\setminus \mathcal{X}_{p_1+1,p_2+1}^{n_1,n_2}$. 
Then, the conditional likelihood function is given by
$$
    L(\bfvartheta \ |\ \mathcal{X}_{1,1}^{n_1,n_2}\setminus \mathcal{X}_{p_1+1,p_2+1}^{n_1,n_2}) := \bbp(\mathcal{X}_{p_1+1,p_2+1}^{n_1,n_2} \ |\ \mathcal{X}_{1,1}^{n_1,n_2}\setminus \mathcal{X}_{p_1+1,p_2+1}^{n_1,n_2})  = \prod^{n_1}_{s=p_1+1} \prod^{n_2}_{t=p_2+1} \bbp(X_{s,t} \,|\, \mathcal{P}\!_{s,t}),
$$
where the conditional probabilities are computed according to Proposition~\ref{prop: CINAR cond distr}.
Here, the factorization of the likelihood function follows from the unilateral model recursion \eqref{eq: CINAR recursion} and the expression \eqref{eq: CINAR cond dist} for the conditional probabilities. 
Then, the log-likelihood function is given by
\begin{equation*}
    \ell(\bfvartheta) = \log L(\bfvartheta) = \sum^{n_1}_{s=p_1+1}\sum^{n_2}_{t=p_2+1} \log \bbp(X_{s,t} \,|\, \mathcal{P}\!_{s,t}).
\end{equation*}
The maximum likelihood estimator is defined as the parameter which maximizes $\ell(\bfvartheta)$ (and thus $L(\bfvartheta)$) and, hence, makes the observed data most ``plausible''. 
In practice, the CML estimates~$\bfvartheta_{\text{CML}}$ are determined by a constrained numerical optimization of the log-likelihood function~$\ell$ (with $\hat{\alpha}_{\text{CML}} = \sum_{(i,j)\in\mathcal{S}} \hat{\theta}_{\text{CML}; ij}$ and $\hat{\phi}_{\text{CML}; ij} = \hat{\theta}_{\text{CML}; ij} / \hat{\alpha}_{\text{CML}}$), where the above YW or CLS estimates can be used as initial values for the numerical routine. Note that it is also possible to numerically approximate the standard errors of the CML estimates (namely from the negative Hessian of~$\ell$ at the maximum, i.e., the observed Fisher information) as well as to compute information criteria for model selection, see \citet[Remark~B.2.1.2]{weiss18} for details. These computations are later illustrated in Section~\ref{Data Example: Yields from an Agricultural Experiment}.

\smallskip
Recall that the conditional distributions of the CINAR random field according to Proposition~\ref{prop: CINAR cond distr} are much simpler than those of the classical INAR random fields in \citet{ghodsi15,sil_wei_26}, as they involve only one convolution. This leads to a clear computational advantage for CML estimation as illustrated for the data example being later discussed in Section~\ref{Data Example: Yields from an Agricultural Experiment} (which is about a rather large data set consisting of large count values as well), always using Poisson innovations. A single evaluation of the respective log-likelihood function takes about 0.4\,s for the CINAR$(1,1)$ model, and about 1.0\,s for the CINAR$(2,2)$ model. For the classical INAR$(1,1)$ model, by contrast, about 54\,s are required for a single evaluation such that a numerical optimization of the log-likelihood function (which requires multiple evaluations) is hardly possible. This is due to the involved three-fold convolutions, and an INAR$(2,2)$ model with its eight-fold convolution is not feasible anyway.

%%%%%%%%%%%%%%%%%%%%%%%%
\section{Performance of Parameter Estimation}
\label{section: Performance of Parameter Estimation}
In order to investigate the finite-sample performance of the estimation approaches derived in Section~\ref{section: Parameter Estimation}, we did a comprehensive simulation study with data generating processes (DGPs) from various types of CINAR models. More precisely, in analogy to \citet{ghodsi12,ghodsi15}, we considered Poi-CINAR$(1,1)$ random fields with $\mu_\varepsilon=1$ and AR parameters $(\theta_{01},\theta_{10},\theta_{11})$ equal to $(0.1,0.1,0.1)$, $(0.2,0.2,0.5)$, or $(0.3,0.4,0.1)$ on the one hand (recall that $\theta_{ij}=\alpha\,\phi_{ij}$), and corresponding NB-counterparts (i.e., with an NB-marginal distribution for~$X_{s,t}$) on the other hand. Here, the required innovations' distribution is computed according to Example~\ref{example: Poi NB marginal} with $\mu_\varepsilon=1$ and $I_\varepsilon:=\sigma_\varepsilon^2/\mu_\varepsilon=2$ throughout. Finally, we also considered the Poi-CINAR$(2,2)$ model with $\mu_\varepsilon=1$ and $(\theta_{01},\theta_{02},\theta_{10},\theta_{11},\theta_{12},\theta_{20},\theta_{21},\theta_{22})$ equal to $(0.1,\ldots,0.1)$ or $(0.15,0.1,0.15,0.15,0.05,0.1,0.05,0.1)$. For each DGP, the sample sizes $(n_1,n_2)$ are chosen as $(10,10)$, $(15,15)$, $(20,20)$, and $(50,50)$, respectively. In order to ensure stationarity, we used a burn-in period of ``width''~100, i.e., we simulated $(n_1+100)\times (n_2+100)$ counts in order to use the $n_1\times n_2$ ``most recent'' counts for parameter estimation. Although our CINAR approach avoids multiple convolutions, the computing time for the constrained optimization via R's \texttt{constrOptim} still quickly increases with increasing model complexity (especially for CML estimation). Therefore, we had to restrict the number of replication to 1,000 per scenario. The full tables with means and standard deviations of the obtained estimates are summarized in Appendix~\ref{appendix: Tabulated Simulation Results}, which are complemented by illustrative boxplots in the subsequent discussion.

\begin{figure}[t]
\centering
$\mu_\varepsilon$\hspace{-2ex}\includegraphics[viewport=30 30 260 375, clip=, scale=0.45]{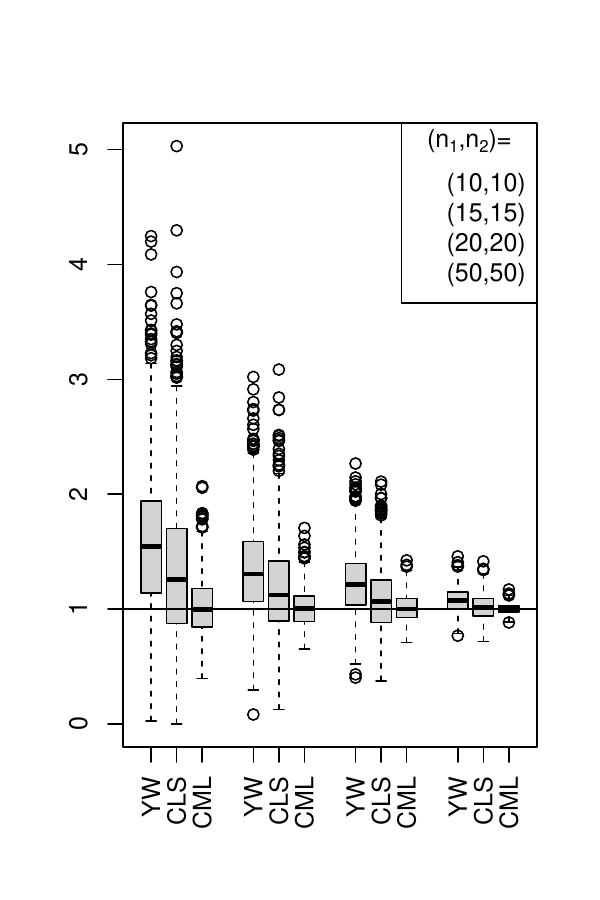}
\quad
$\theta_{01}$\hspace{-2ex}\includegraphics[viewport=30 30 260 375, clip=, scale=0.45]{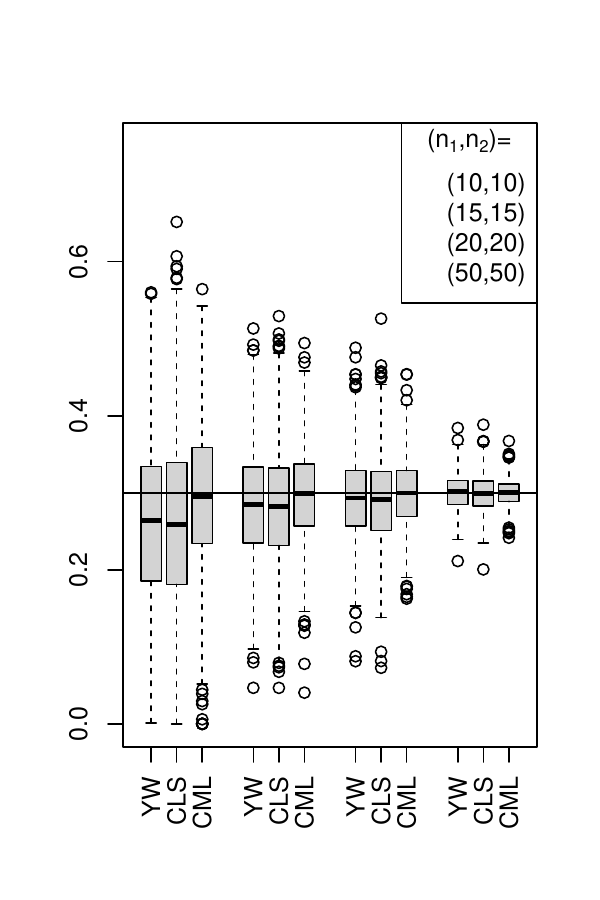}
\quad
$\theta_{10}$\hspace{-2ex}\includegraphics[viewport=30 30 260 375, clip=, scale=0.45]{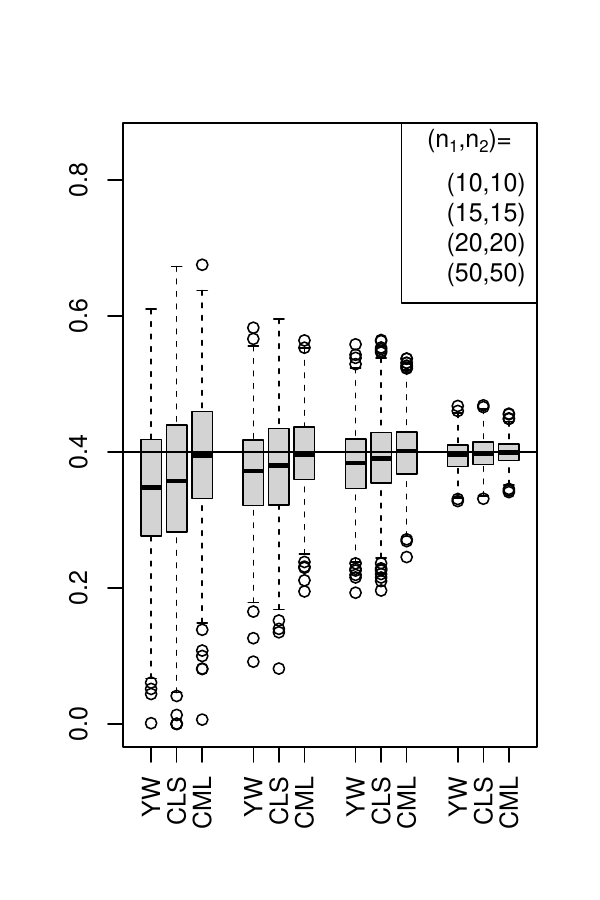}
\quad
$\theta_{11}$\hspace{-2ex}\includegraphics[viewport=30 30 260 375, clip=, scale=0.45]{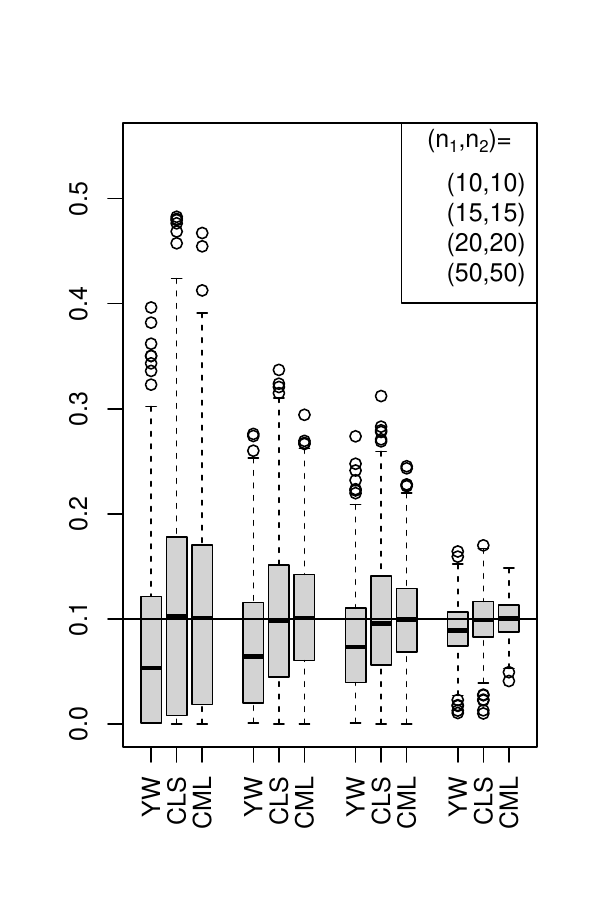}
\caption{Boxplots of simulated estimates for Poi-CINAR$(1,1)$ random field (true parameter values at horizontal lines), with increasing sample size $(n_1,n_2)$ from left to right.}
\label{fig: poicinar11 boxplots}
\end{figure}

\smallskip
Let us start our analyses with the Poi-CINAR$(1,1)$ random field. As can be seen from Table~\ref{tab_sim_poicinar11} in the appendix, in the setting $(\theta_{01},\theta_{10},\theta_{11}) = (0.1,0.1,0.1)$ with low AR parameters, all three estimation methods work similarly well. This clearly changes, however, in the remaining two scenarios, where at least one of the parameters is rather large. Then, CLS and in particular YW estimation lead to a strong negative bias for the large dependence parameters, which, in turn, causes a strong positive bias for~$\mu_\varepsilon$. The CML estimator, by contrast, shows only negligible bias throughout, and also its standard deviations are substantially lower than those of YW and CLS. This discrepancy in performance can also be seen in Figure~\ref{fig: poicinar11 boxplots}, where the boxplots for the scenario $(\theta_{01},\theta_{10},\theta_{11}) = (0.3,0.4,0.1)$ are shown. Altogether, we conclude that the Poi-CINAR$(1,1)$'s parameters should be estimated with the CML~approach, whereas YW and CLS are only recommended for initializing the required numerical optimization routine.

\begin{figure}[t]
\centering
$\mu_\varepsilon$\hspace{-2ex}\includegraphics[viewport=30 15 335 375, clip=, scale=0.45]{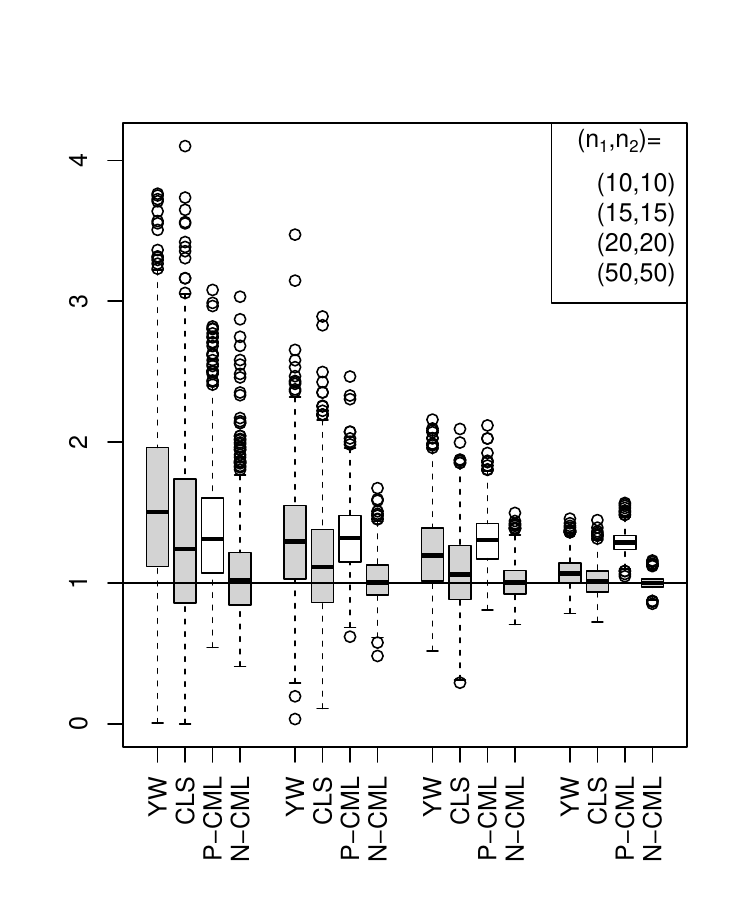}
\quad
$I_\varepsilon$\hspace{-2ex}\includegraphics[viewport=30 15 225 375, clip=, scale=0.45]{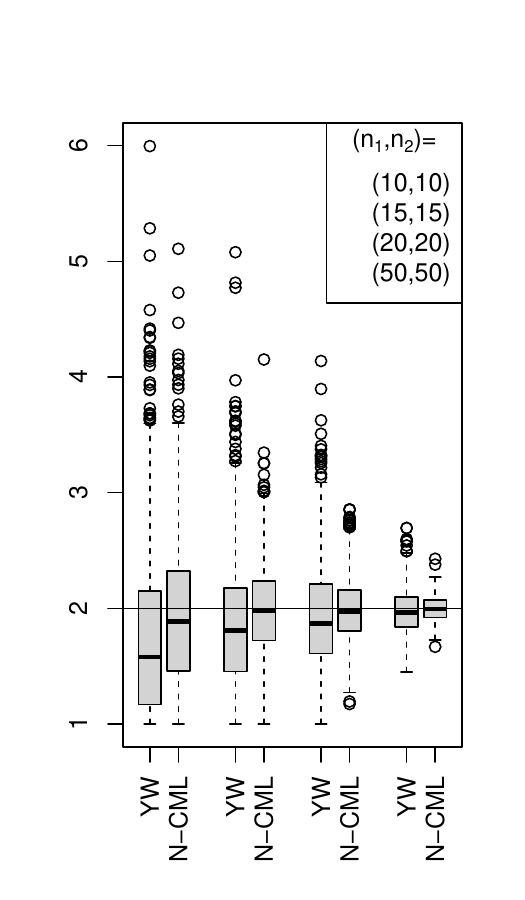}
\\[2ex]
$\theta_{01}$\hspace{-2ex}\includegraphics[viewport=30 15 335 375, clip=, scale=0.45]{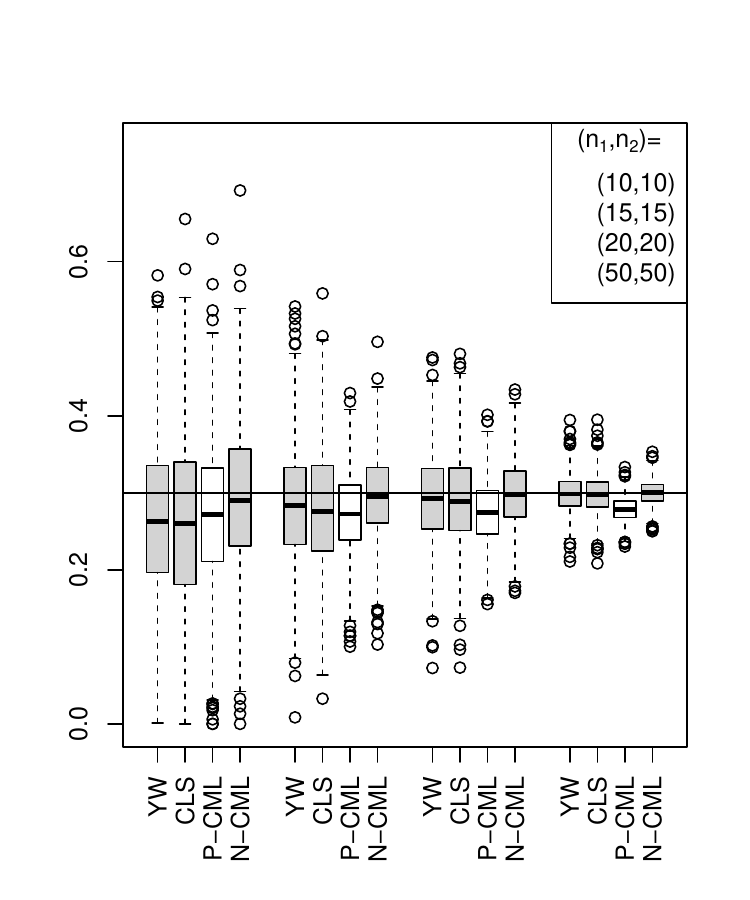}
\quad
$\theta_{10}$\hspace{-2ex}\includegraphics[viewport=30 15 335 375, clip=, scale=0.45]{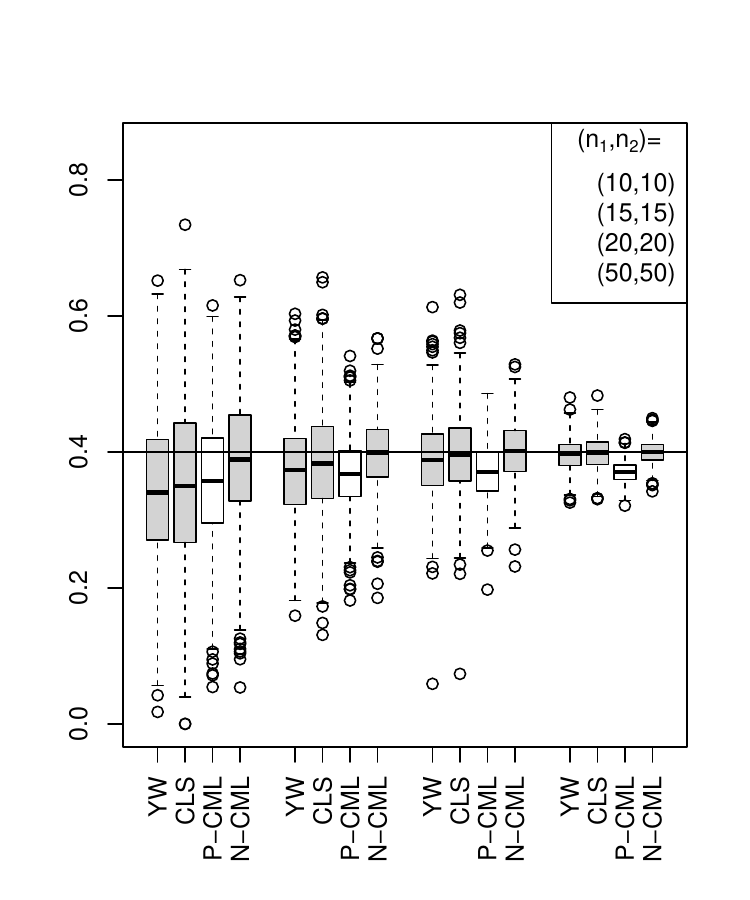}
\quad
$\theta_{11}$\hspace{-2ex}\includegraphics[viewport=30 15 335 375, clip=, scale=0.45]{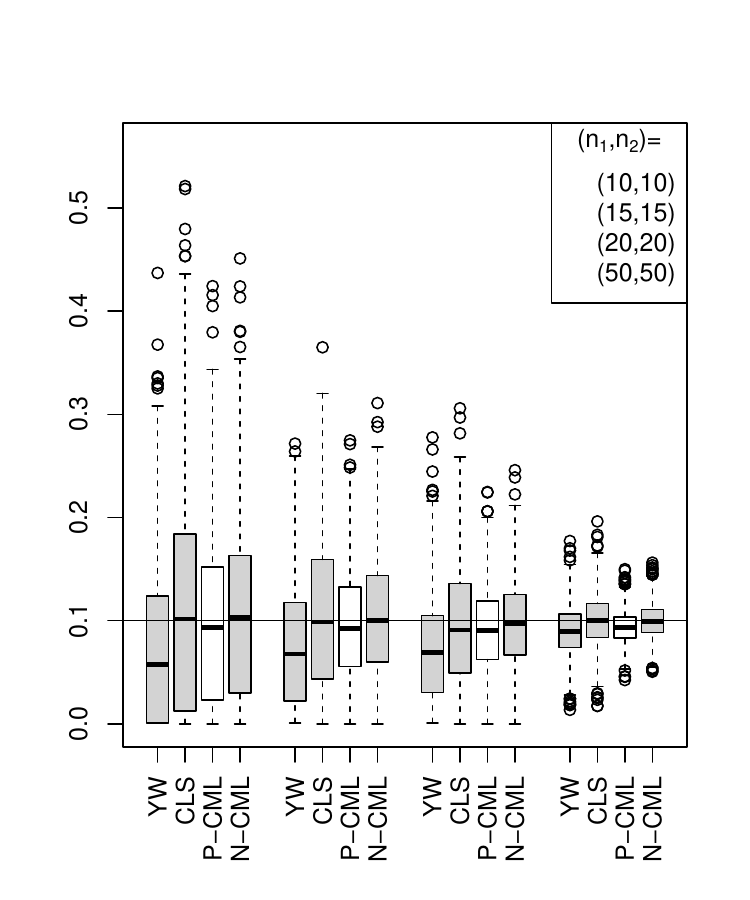}
\caption{Boxplots of simulated estimates for NB-CINAR$(1,1)$ random field (true parameter values at horizontal lines), with increasing sample size $(n_1,n_2)$ from left to right. Boxplots with white background correspond to model misspecification.}
\label{fig: nbcinar11 boxplots}
\end{figure}

\smallskip
Next, we consider the NB-CINAR$(1,1)$ DGPs, the innovations of which uniquely have the dispersion ratio $I_\varepsilon=2$, i.e., they exhibit a substantial extent of overdispersion. Recall that CLS is only able to estimate the parameters being involved in the conditional mean, i.e., it cannot estimate the value of~$I_\varepsilon$. The latter is only achieved by YW and CML (more precisely, by ``N-CML'', see the discussion below). The results in Table~\ref{tab_sim_nbcinar11} and Figure~\ref{fig: nbcinar11 boxplots} confirm our previous conclusion that YW and CLS are only competitive to CML if the AR parameters are low. Otherwise, we have the same problems with bias and standard deviation described before, with the YW~estimator of~$I_\varepsilon$ exhibiting a strong negative bias. So our previous recommendation of using CML for parameter estimation extends to the NB-case as well. However, there is one point to be noted. The outstanding performance of CML estimation clearly depends on choosing the adequate model type. To analyze the effect of a possible model misspecification, we did not only use the NB-CINAR$(1,1)$ likelihood function for parameter estimation (labeled as ``N-CML''), but also falsely the one of a Poi-CINAR$(1,1)$ random field (labeled as ``P-CML''). In Figure~\ref{fig: nbcinar11 boxplots}, the misspecified case is further highlighted by white boxes. It gets clear that under model misspecification, the CML estimates are affected by a strong bias that also does not vanish with increasing sample size. Then, for large sample sizes, even YW and CLS show a better performance than CML. So while CML is generally recommended due to its outstanding performance, it should always be accompanied by subsequent checks of model adequacy. A substantial discrepancy between CLS and CML, in turn, should be interpreted as a warning sign of a possible model misspecification.

\begin{figure}[th!]
\centering
$\mu_\varepsilon$\hspace{-2ex}\includegraphics[viewport=30 15 335 375, clip=, scale=0.45]{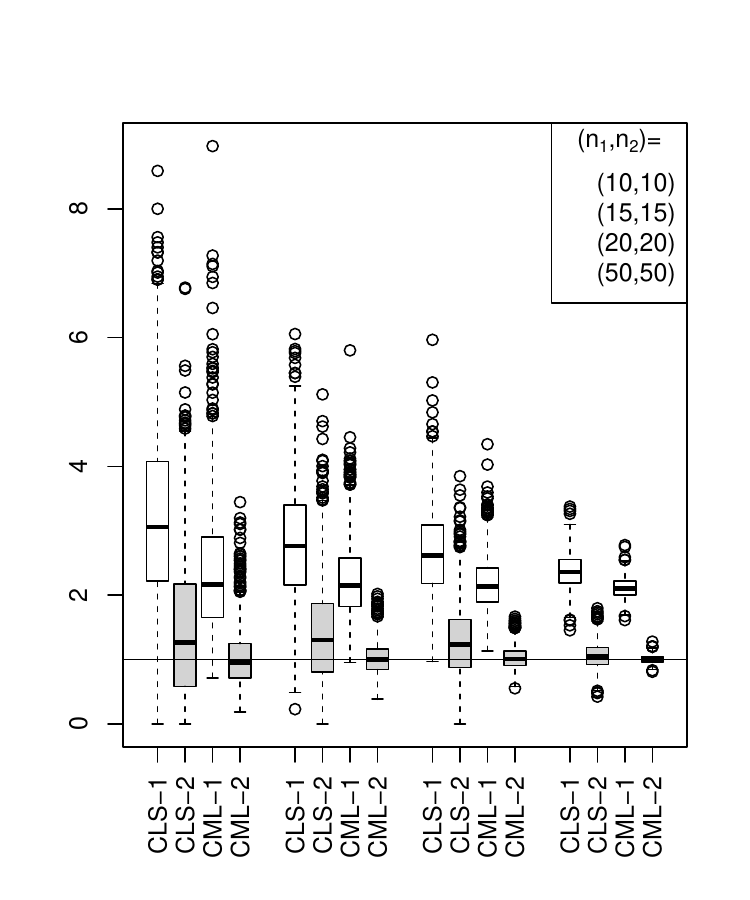}
\quad
$\theta_{01}$\hspace{-2ex}\includegraphics[viewport=30 15 335 375, clip=, scale=0.45]{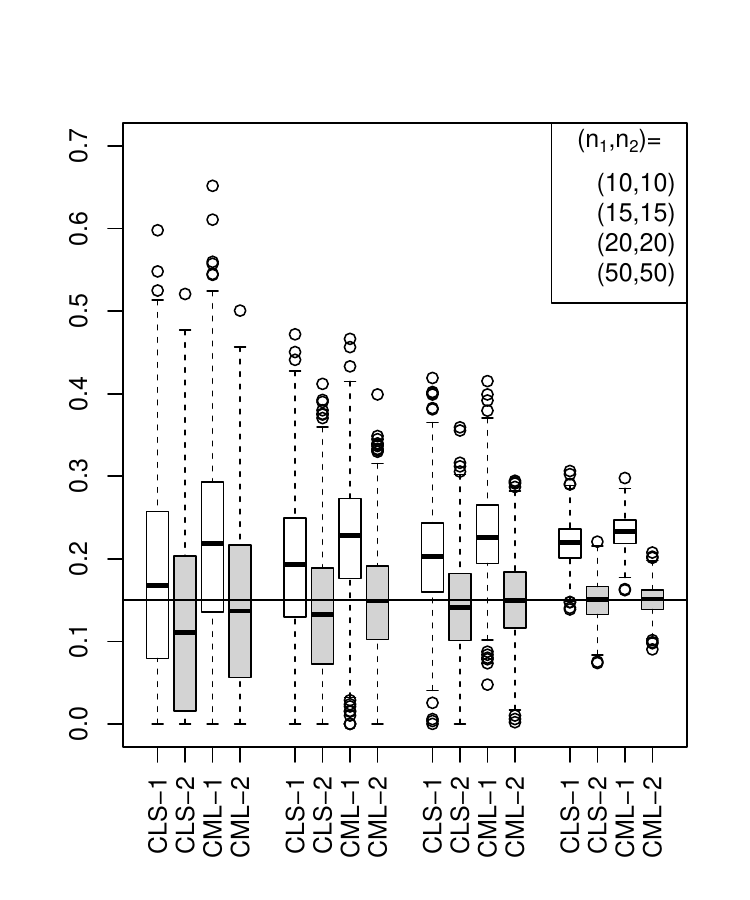}
\quad
$\theta_{02}$\hspace{-2ex}\includegraphics[viewport=30 15 225 375, clip=, scale=0.45]{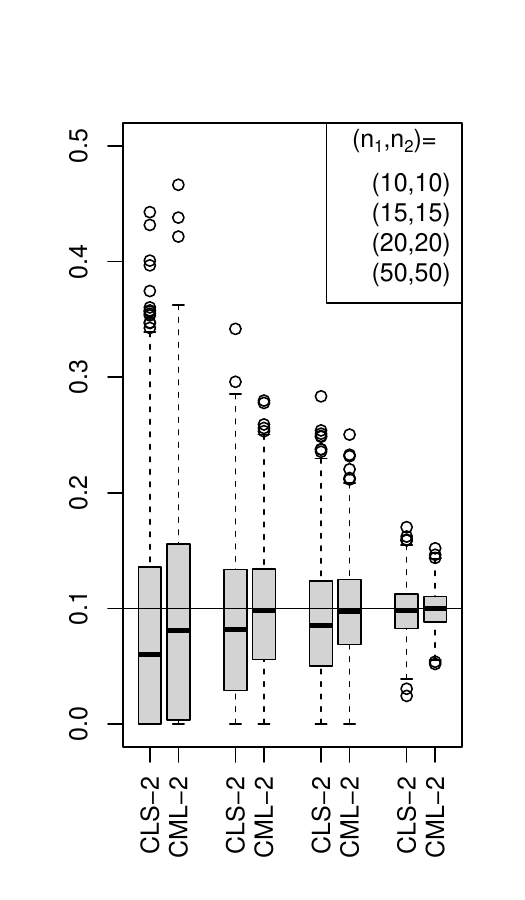}
\\[2ex]
$\theta_{10}$\hspace{-2ex}\includegraphics[viewport=30 15 335 375, clip=, scale=0.45]{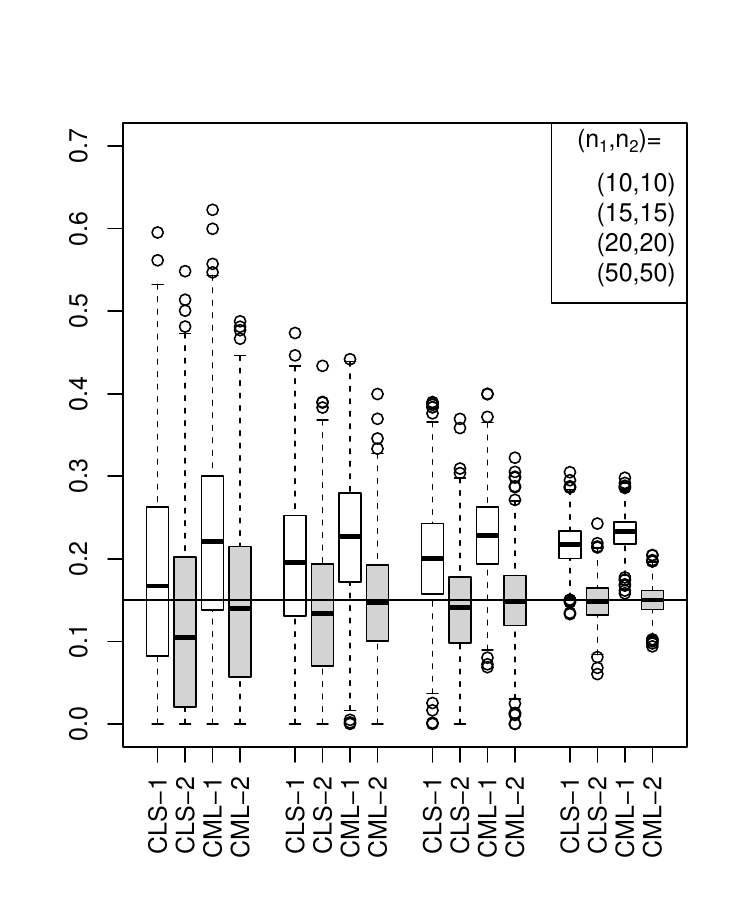}
\quad
$\theta_{11}$\hspace{-2ex}\includegraphics[viewport=30 15 335 375, clip=, scale=0.45]{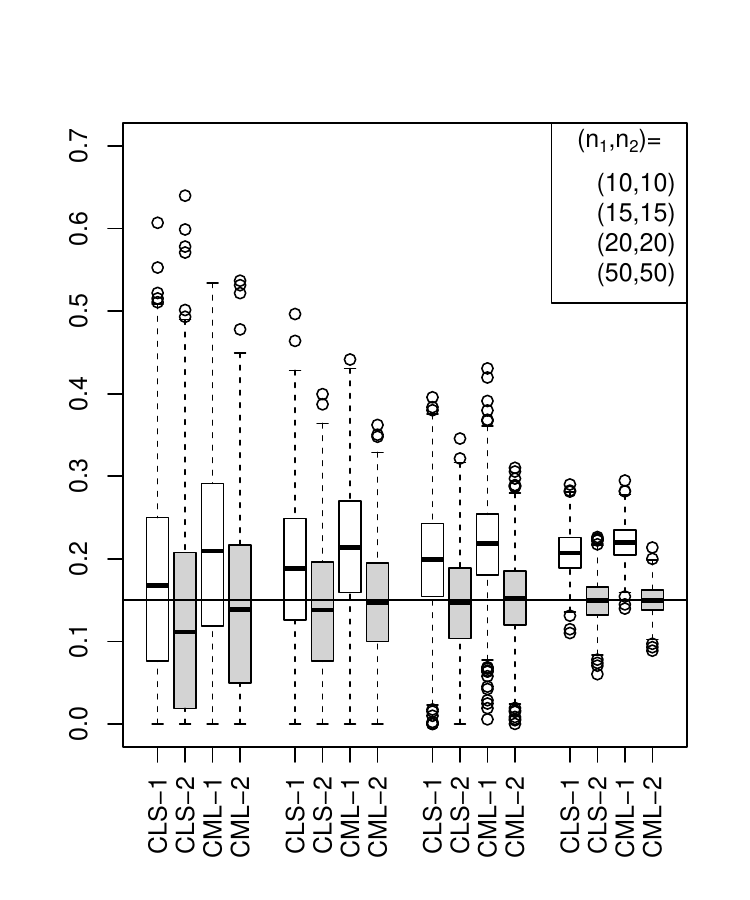}
\quad
$\theta_{12}$\hspace{-2ex}\includegraphics[viewport=30 15 225 375, clip=, scale=0.45]{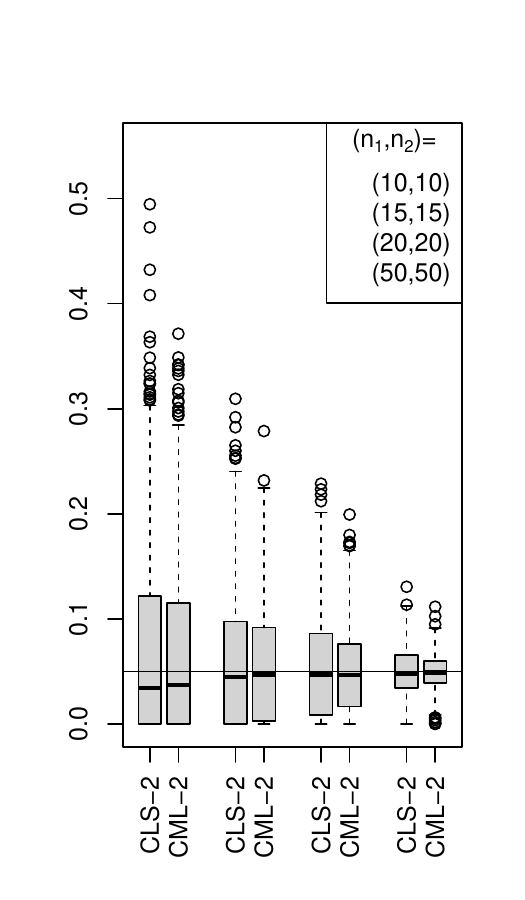}
\\[2ex]
$\theta_{20}$\hspace{-2ex}\includegraphics[viewport=30 15 225 375, clip=, scale=0.45]{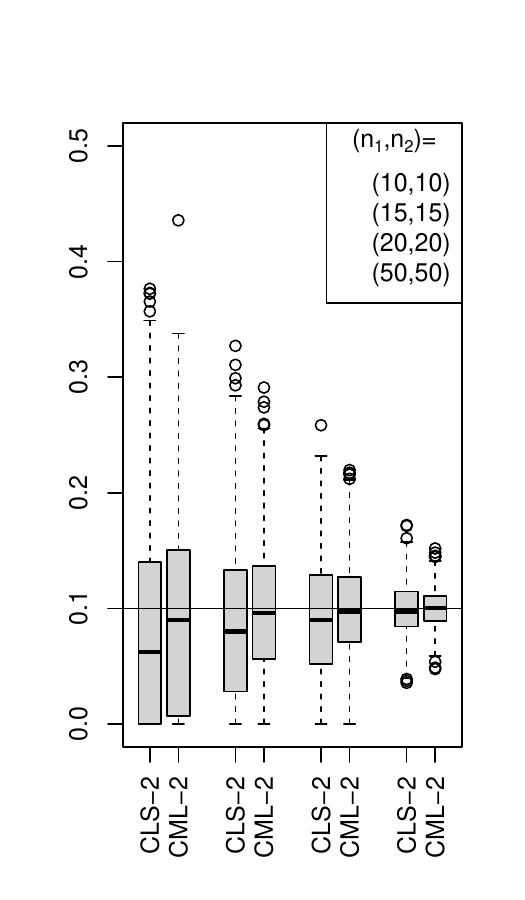}
\quad
$\theta_{21}$\hspace{-2ex}\includegraphics[viewport=30 15 225 375, clip=, scale=0.45]{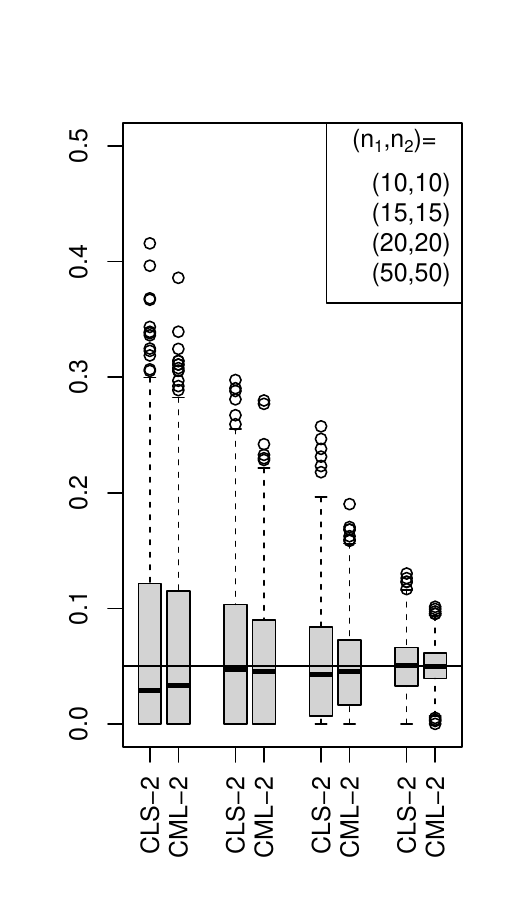}
\quad
$\theta_{22}$\hspace{-2ex}\includegraphics[viewport=30 15 225 375, clip=, scale=0.45]{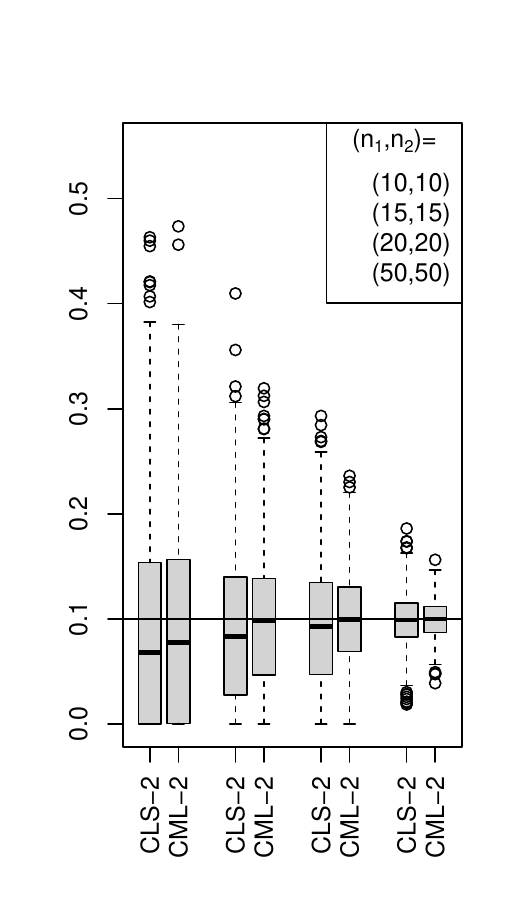}
\caption{Boxplots of simulated estimates for Poi-CINAR$(2,2)$ random field (true parameter values at horizontal lines), with increasing sample size $(n_1,n_2)$ from left to right. Boxplots with white background correspond to model misspecification.}
\label{fig: poicinar22 boxplots}
\end{figure}

\smallskip
Finally, we turn to the higher-order case, namely the aforementioned Poi-CINAR$(2,2)$ DGPs, see Table~\ref{tab_sim_poicinar22} and Figure~\ref{fig: poicinar22 boxplots} for a summary. Here, we did not use YW anymore as it becomes more and more difficult to obtain admissible estimates (i.e., satisfying all parameter constraints) with increasing number of parameters. On the other hand, we extended our analyses of a possible model misspecification by considering both the CLS and CML estimates corresponding to falsely fitting a Poi-CINAR$(1,1)$ model (labeled as ``CLS-1'' and ``CML-1'', respectively, and again highlighted by white boxes in Figure~\ref{fig: poicinar22 boxplots}). The results obtained are consistent with our previous observations. For a correct model specification, CLS improves with increasing sample size, but is clearly outperformed by CML in terms of both bias and standard deviation. So using CLS for initial and CML for final parameter estimation is still recommended. If the model order is misspecified, by contrast, both estimation approaches show a poor performance and a non-vanishing bias also for large sample sizes. So again, we advise to not do model fitting without a subsequent model diagnosis.

\begin{figure}[t]
\centering
% (a)\hspace{-3ex}\includegraphics[viewport=30 45 630 240, clip=, scale=0.75]{wheat_plot_old.pdf}
(a)\includegraphics[viewport=30 55 630 260, clip=, scale=0.70]{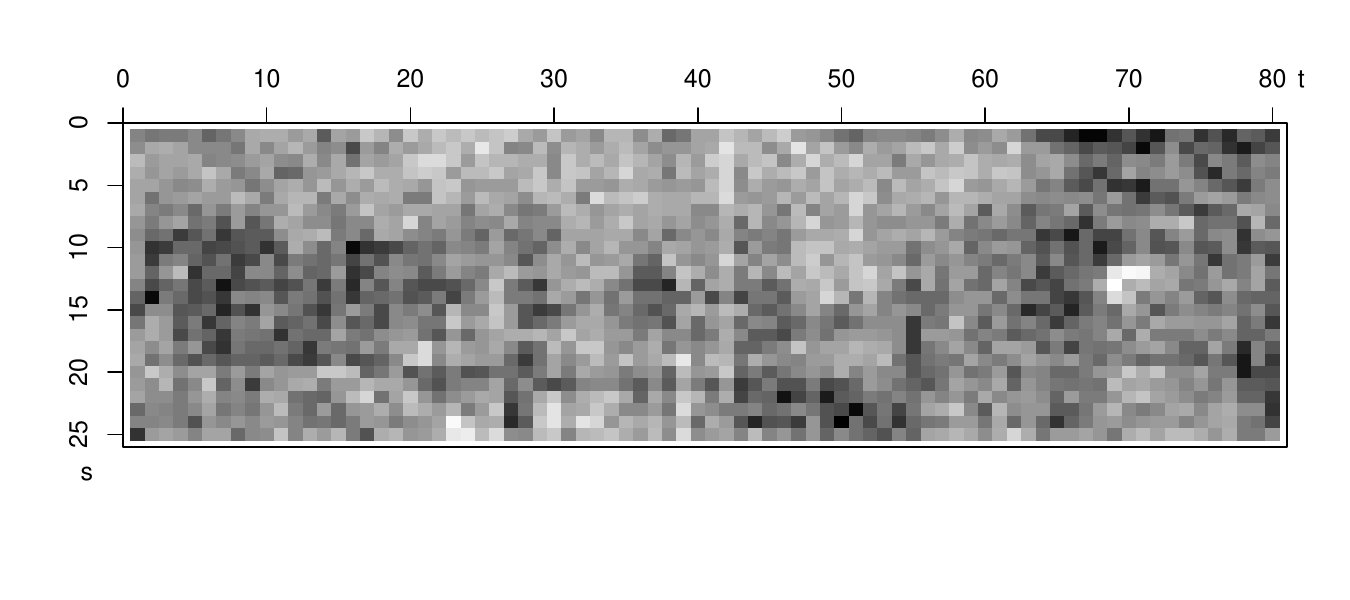}
\\[2ex]
\begin{tabular}{@{}l@{}}
(b)\hspace{-3ex}\includegraphics[viewport=0 45 330 230, clip=, scale=0.7]{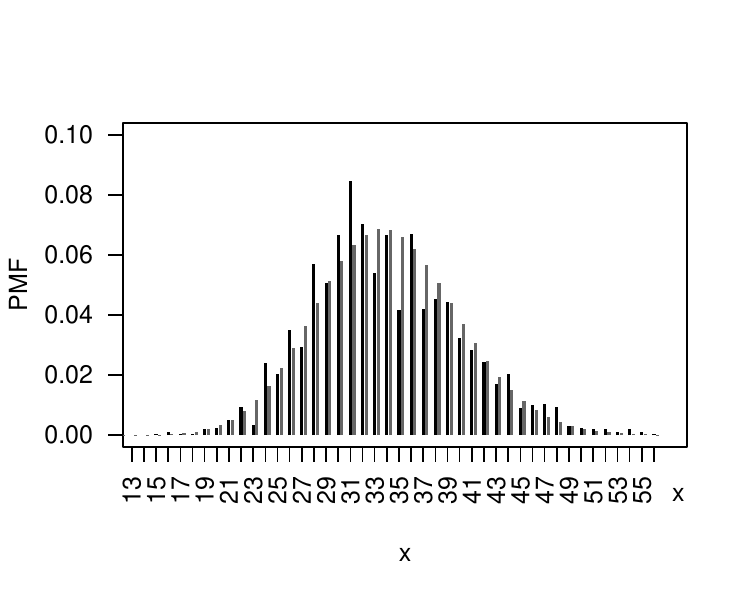}
\end{tabular}\qquad
\begin{tabular}{@{}l@{}}
(c)\quad Sample ACF at lag $(k,l)$:
\\[1ex]
$\begin{array}{rccccc}
\toprule
l\ \setminus\ k & -2 & -1 & 0 & 1 & 2 \\
\midrule
2 & 0.228 & 0.339 & 0.422 & 0.302 & 0.231 \\
1 & 0.254 & 0.369 & 0.518 & 0.352 & 0.240 \\
0 & 0.349 & 0.511 & 1.000 & 0.511 & 0.349 \\
-1 & 0.240 & 0.352 & 0.518 & 0.369 & 0.254 \\
-2 & 0.231 & 0.302 & 0.422 & 0.339 & 0.228 \\
\bottomrule
\end{array}$
\end{tabular}
\caption{Wheat yields data from Section~\ref{Data Example: Yields from an Agricultural Experiment}: (a) gray-scale plot of data, where counts range
between 0 (white) and 56 (black), (b) sample PMF (black) compared to Poi(33.8425)-PMF
(gray), and (c) sample ACF for spatial lags $(k,l)$ with $-2\leq k,l\leq 2$.}
\label{fig: wheat}
\end{figure}

\section{Data Example: Yields from an Agricultural Experiment}
\label{Data Example: Yields from an Agricultural Experiment}

We consider the data set published in the Appendix of \citet{iyer42}, which is also offered by the R~package ``\href{https://cran.r-project.org/package=agridat}{agridat}'' through the command \texttt{iyer.wheat.uniformity}. The data are plotted in Figure~\ref{fig: wheat}\,(a), in the same arrangement as in the table in \citet{iyer42}. They describe the result of an agricultural experiment about harvesting wheat in an experimental field that was divided into $n_1\times n_2=25\times 80$ plots of size $5\,\textup{ft}\times 5\,\textup{ft}$, with the grain yield~$X_{s,t}$ reported in units of half an ounce. The resulting sample PMF in Figure~\ref{fig: wheat}\,(b) is quite similar to a Poi-distribution with parameter equal to the sample mean $\hat\mu_X=33.8425$. The close agreement to a Poi-distribution is also confirmed by the sample dispersion ratio $\hat\sigma_X^2/\hat\mu_X\approx 1.129$, which shows that the data are nearly equidispersed.
%var 38.2088
Finally, the sample ACF in part~(c) shows substantially positive ACF values that quickly decay with increasing absolute spatial lag such that altogether, it appears reasonable to try to model the data by an AR-type model having a Poi-marginal distribution, as it is the case for the Poi-CINAR model.

\begin{table}[t]
\centering
\caption{Wheat yields data from Section~\ref{Data Example: Yields from an Agricultural Experiment}: CML estimates (standard errors in parentheses) and information criteria for CINAR$(1,1)$ candidate models.}
\label{tab: clm cinar11}

\smallskip
\begin{tabular}{lrrrrrr}
\toprule
Model & $\mu_\varepsilon$ & $\theta_{01}$ & $\theta_{10}$ & $\theta_{11}$ & AIC & BIC \\
\midrule
full & 11.215 & 0.287 & 0.358 & 0.023 & 11932.9 & 11955.3 \\
 & \footnotesize (0.474) & \footnotesize (0.022) & \footnotesize (0.022) & \footnotesize (0.021) &  &  \\
\midrule
simplified & 11.428 & 0.296 & 0.365 & 0 & 11932.2 & 11949.0 \\
 & \footnotesize (0.439) & \footnotesize (0.021) & \footnotesize (0.021) &  &  &  \\
\bottomrule
\end{tabular}
\end{table}

\smallskip
We start our analyses with the (full) Poi-CINAR$(1,1)$ model according to Section~\ref{subsection: Special Case: The CINAR(1,1) Model}. The corresponding CML estimates are determined by a numerical optimization of the log-likelihood function~$\ell$, where the YW or CLS estimates can be used as initial values for the numerical routine. Here, in view of a unique parametrization across the different estimation approaches, we use the parametrization with $\theta_{ij}=\alpha\,\phi_{ij}$. Approximate standard errors for the CML estimates are computed from the negative Hessian of~$\ell$ at the maximum (observed Fisher information), see Remark~B.2.1.2 in \citet{weiss18} for details. The results are summarized in the first line of Table~\ref{tab: clm cinar11}. If judging the estimates on a 5\,\%-level, it gets clear that the estimate~$\theta_{11}$ (expressing a diagonal dependence) is not significantly different from zero, which implies that a simplified CINAR$(1,1)$ model with $\theta_{11}\equiv 0$ (so with spatial dependence only in horizontal and vertical direction) appears to be another reasonable candidate model; see Table~\ref{tab: clm cinar11} for the corresponding model fit. In order to choose among the different candidate models, a common approach is to use information criteria such as Akaike's or the Bayesian one (AIC or BIC, respectively). Since these criteria are computed from the \emph{conditional} log-likelihood and since we shall later also compare to different model orders, we use a re-scaled maximized log-likelihood~$\ell_{\max}$ like in \citet[Remark~B.2.1.2]{weiss18} for AIC and BIC computation, namely $n_1n_2/n_\ell\cdot \ell_{\max}$ where $n_\ell$ expresses the number of summands in the conditional log-likelihood function. From the results in Table~\ref{tab: clm cinar11}, we recognize that both criteria prefer the simplified CINAR$(1,1)$ model. 

\begin{figure}[t]
\centering
\begin{tabular}{@{}l@{}}
(a)\quad \textbf{Standardized Pearson residuals:}\\
mean $-0.004$, variance $0.983$,\\
and sample ACF at lag $(k,l)$:
\\[1ex]
$\begin{array}{rccccc}
\toprule
l\ \setminus\ k & -2 & -1 & 0 & 1 & 2 \\
\midrule
2 & 0.025 & 0.067 & 0.119 & 0.059 & 0.058 \\
1 & 0.025 & 0.013 & 0.022 & 0.074 & -0.010 \\
0 & 0.083 & 0.007 & 1.000 & 0.007 & 0.083 \\
-1 & -0.010 & 0.074 & 0.022 & 0.013 & 0.025 \\
-2 & 0.058 & 0.059 & 0.119 & 0.067 & 0.025 \\
\bottomrule
\end{array}$
\end{tabular}
\qquad
\begin{tabular}{@{}l@{}}
(b)\hspace{-3ex}\includegraphics[viewport=0 45 345 235, clip=, scale=0.65]{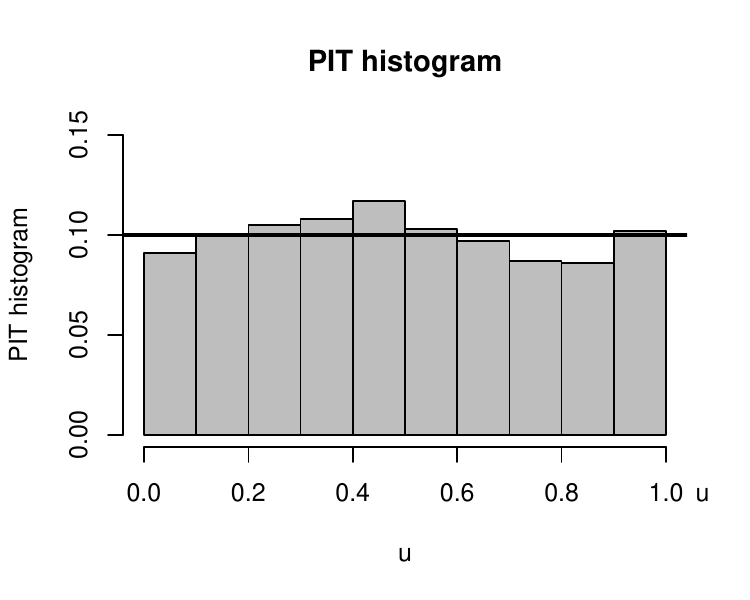}
\end{tabular}
\caption{Simplified CINAR$(1,1)$ model for wheat yields data according to Table~\ref{tab: clm cinar11}: (a) mean, variance, and sample ACF of Pearson residuals; (b) PIT histogram.}
\label{fig: wheat cinar11}
\end{figure}

\smallskip
In order to check the adequacy of the selected model, we use the standardized Pearson residuals and the PIT histogram for model diagnostics (computed according to Proposition~\ref{prop: CINAR cond distr}), which are established tools in the analysis of count time series \citep[Section~2.4]{weiss18} as well as spatial count data \citep{jung22}. The mean and variance of the Pearson residuals in Figure~\ref{fig: wheat cinar11} are rather close to zero and one, respectively, which is in favor of model validity. Also the PIT histogram is reasonably close to uniformity and thus confirms an adequate model choice. If comparing the sample ACF of the Pearson residuals to those of the original data, recall Figure~\ref{fig: wheat}\,(c), we recognize a strong reduction of absolute value, i.e., most part of the apparent dependencies are covered by the simplified CINAR$(1,1)$ model. A closer look, however, still reveals relatively large values at lags $(0,\mp 2)$ and $(\mp 2,0)$, which indicates that a second-order CINAR model might be better suited for the wheat yields data.

\begin{table}[t]
\centering
\caption{Wheat yields data from Section~\ref{Data Example: Yields from an Agricultural Experiment}: CML estimates (standard errors in parentheses) and information criteria for CINAR$(2,2)$ candidate models.}
\label{tab: clm cinar22}

\smallskip
\begin{tabular}{lrrrrrrrrrrr}
\toprule
Model & $\mu_\varepsilon$ & $\theta_{02}$ & $\theta_{01}$ & $\theta_{10}$ & $\theta_{11}$ & $\theta_{12}$ & $\theta_{20}$ & $\theta_{21}$ & $\theta_{22}$ & AIC & BIC \\
\midrule
full & 9.420 & 0.233 & 0.100 & 0.313 & 0.002 & 0.016 & 0.057 & 0.000 & 0.000 & 11868.0 & 11918.4 \\
 & \footnotesize (0.542) & \footnotesize (0.025) & \footnotesize (0.023) & \footnotesize (0.026) & \footnotesize (0.024) & \footnotesize (0.020) & \footnotesize (0.023) & \footnotesize (0.015) & \footnotesize (0.021) &  &  \\
simplified & 9.498 & 0.238 & 0.104 & 0.313 & 0 & 0 & 0.065 & 0 & 0 & 11860.2 & 11888.2 \\
 & \footnotesize (0.481) & \footnotesize (0.023) & \footnotesize (0.022) & \footnotesize (0.025) &  &  & \footnotesize (0.023) &  &  &  &  \\
\bottomrule
\end{tabular}
\end{table}

\begin{figure}[t]
\centering
\begin{tabular}{@{}l@{}}
(a)\quad \textbf{Standardized Pearson residuals:}\\
mean $-0.004$, variance $0.935$,\\
and sample ACF at lag $(k,l)$:
\\[1ex]
$\begin{array}{rccccc}
\toprule
l\ \setminus\ k & -2 & -1 & 0 & 1 & 2 \\
\midrule
2 & -0.002 & 0.054 & 0.032 & 0.048  & 0.040 \\
1 & 0.025 & 0.018 & 0.067 & 0.074 & -0.006 \\
0 & 0.032 & 0.030 & 1.000 & 0.030 & 0.032 \\
-1 & -0.006 & 0.074 & 0.067 & 0.018 & 0.025 \\
-2 & 0.040 & 0.048 & 0.032 & 0.054 & -0.002 \\
\bottomrule
\end{array}$
\end{tabular}
\qquad
\begin{tabular}{@{}l@{}}
(b)\hspace{-3ex}\includegraphics[viewport=0 45 345 235, clip=, scale=0.65]{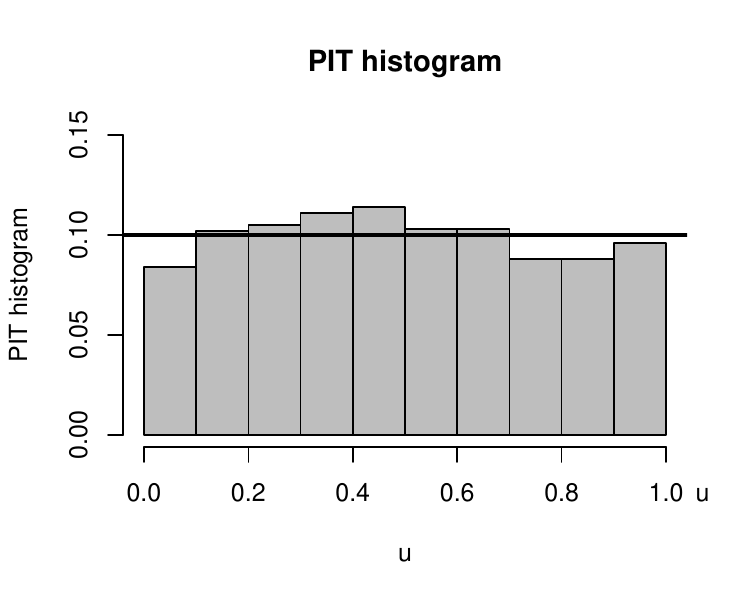}
\end{tabular}
\caption{Simplified CINAR$(2,2)$ model for wheat yields data according to Table~\ref{tab: clm cinar22}: (a) mean, variance, and sample ACF of Pearson residuals; (b) PIT histogram.}
\label{fig: wheat cinar22}
\end{figure}

\smallskip
Hence, we first fitted a full CINAR$(2,2)$ model to the wheat yields data, see the first line in Table~\ref{tab: clm cinar22}. We recognize that only the estimates for $\theta_{02},\theta_{01}$ (vertical dependence) and $\theta_{10},\theta_{20}$ (horizontal dependence) are significantly different from zero. These results nicely agree with our first-order findings, and it extends the dependence structure in those directions where relatively large values of the CINAR$(1,1)$ residuals' ACF were detected, recall Figure~\ref{fig: wheat cinar22}\,(a). Table~\ref{tab: clm cinar22} also shows the estimates for the simplified CINAR$(2,2)$ model. From the given AIC and BIC values, we conclude that this model does not only outperform the full CINAR$(2,2)$ model, but also any of the first-order models from Table~\ref{tab: clm cinar11}. Finally, let us look at the correspondings model diagnostics in Figure~\ref{fig: wheat cinar22}. While PIT~histogram and the Pearson residuals' mean and variance indicate a similar performance as for the simplified CINAR$(1,1)$ model, we recognize a clear improvement in terms of their ACF, where all values deviate from zero only in the second decimal place. Although none of the criteria in Figure~\ref{fig: wheat cinar22} indicates a perfect adequacy, it seems that the simplified CINAR$(2,2)$ model constitutes a reasonable ``workhorse model'' for the wheat yields data.

% %%%%%%%%%%%%%%%%%%%%%%%%
\section{Conclusions and Future Research}
\label{section: Conclusions and Future Research}
In this article, we proposed the CINAR family for modeling count random fields, which exhibits the same autocorrelation structure as ordinary INAR random fields, but which can be equipped with, e.g., a Poi- or NB-marginal distribution. More generally, having specified the innovations' distribution, the observations' marginal distribution is fully known, which differs from the case of INAR random fields. A further advantage is a simpler computation of conditional probabilities and, thus, likelihood functions. In fact, we developed three approaches for parameter estimation (YW, CLS, and CML) and analyzed their performance by simulations. If the model is correctly specified, the CML approach clearly surpasses its competitors, but it performs poorly under model misspecification. So model fitting should not be done without subsequent checks of model adequacy. This was exemplified by our data example on wheat yields from an agricultural experiment, which also demonstrated the practical relevance of our novel CINAR family.

\smallskip
There are various directions for future research. First, one may adapt the CINAR model to different INAR-like models for counts, e.g., to the iterated-thinning INAR model having both NB-distributed observations and innovations, see \citet{aleksandrov24} for a comprehensive discussion. Second, one could develop a CINAR-like model for bounded counts (i.e., where the range is bounded from above by some integer value $N\in\bbn$), where the binomial AR model of \citet{weiss09} appears to be a possible starting point. Finally, as already discussed in some detail in Section~\ref{subsection: Outlook on Possible Extensions}, a multilateral extension of the CINAR model as well as a Tobit approach for negative dependencies appear to be appealing challenges for future research.

%%%%%%%%%%%%%%%%%%%%%%
%\subsubsection*{Acknowledgments}
%The authors thank the associate editor and the referees for their useful comments on an earlier draft of this article.

\bibliographystyle{plainnat}
\bibliography{references}
\vspace{12pt}

%\newpage
\appendix
\numberwithin{table}{section}

%to avoid "Appendix" in references caused by elsarticle:
\gdef\thesection{\Alph{section}} % corrected redefinition of "\thesection"
\makeatletter
\renewcommand\@seccntformat[1]{\appendixname\ \csname the#1\endcsname.\hspace{0.5em}}
\makeatother

\section{Derivations}
\label{appendix: Derivations}

\subsection{Proof of Proposition~\ref{prop: ergodicity}}
\label{Proof of Proposition prop: ergodicity}
Let $\mathbf{Z}(s,t)$ denote all Bernoulli counting series involved in $X_{s,t}$ at ``time'' $(s,t)$. From the model recursion in \eqref{eq: CINAR recursion}, we see that $X_{s,t}$ can be understood as a function
%, say $f$,
of $(\mathbf{D}_{s,t}, \varepsilon_{s,t}, \mathbf{Z}(s,t))$. Hence, we may write
\[
	\sigma(X_{s,t}, X_{s-1,t}, X_{s,t-1}, \dots) \subseteq \sigma\Big(\mathbf{D}_{s,t}, \varepsilon_{s,t}, \mathbf{Z}(s,t), \mathbf{D}_{s-1,t}, \varepsilon_{s-1,t}, \mathbf{Z}(s-1,t), \dots\Big),
\]
where $\sigma(Y)$ denotes the $\sigma$-field generated by $Y$, and 
\[
	\bigcap^\infty_{s=1} \bigcap^\infty_{t=1} \sigma(X_{s,t}, X_{s-1,t}, \dots) \subseteq \bigcap^\infty_{s=1} \bigcap^\infty_{t=1} \sigma\Big(\mathbf{D}_{s,t}, \varepsilon_{s,t}, \mathbf{Z}(s,t), \mathbf{D}_{s-1,t}, \varepsilon_{s-1,t}, \mathbf{Z}(s-1,t), \dots\Big).
\]
The sequences $(\mathbf{D}_{s,t}), (\varepsilon_{s,t})$ and $(\mathbf{Z}(s,t))$ are i.i.d.\ as well as independent of each other by assumption, so $(\mathbf{D}_{s,t}, \varepsilon_{s,t}, \mathbf{Z}(s,t))$ is i.i.d.\ itself. Kolmogorov's zero-one law (see, e.g., \citealp[Theorem~22.3]{billingsley95}) gives that all events in this tail $\sigma$-field are trivial, i.e., every set in the tail $\sigma$-field has probability $0$ or $1$. Accordingly, the random field model $(X_{s,t})$ is ergodic and the proof of Proposition~\ref{prop: ergodicity} is complete.

\subsection{Proof of Proposition~\ref{prop: CINAR marginal properties}}
\label{Proof of Proposition prop: CINAR marginal properties}
If $(X_{s,t})$ is stationary, then the marginal pgf is given by 
\begin{align*}
    \pgf_X(u) &= \bbe(u^{X_{s,t}}) = \bbe(\bbe(u^{X_{s,t}}\, |\, \mathbf{D}_{s,t})) = \sum_{(i,j)\in \mathcal{S}} \phi_{ij}\cdot \bbe(u^{\alpha\circ X_{s-i,t-j}+\varepsilon_{s,t}}) \nonumber \\
    &= \sum_{(i,j)\in \mathcal{S}} \phi_{ij}\cdot\pgf_X(1-\alpha+\alpha u) \cdot\pgf_\varepsilon(u) = \pgf_X(1-\alpha+\alpha u) \cdot \pgf_\varepsilon(u), 
\end{align*}
as $(1-\pi+\pi u)^n$ is the pgf of a binomially distributed random variable with parameters $n\in\bbn$ and $\pi\in(0,1)$, and $\sum_{(i,j)\in\mathcal{S}} \phi_{ij}=1$.
Since the pgf captures the marginal distribution of the model and is identical to the pgf of the classical INAR$(1)$ time series model, it follows that the CINAR's mean and variance must also agree with those of the INAR$(1)$ model. So the proof of Proposition~\ref{prop: CINAR marginal properties} is complete.

\subsection{Proof of Proposition~\ref{prop: CINAR spatial properties}}
\label{Proof of Proposition prop: CINAR spatial properties}
In order to derive the autocovariance function (ACvF) of the CINAR random field, we first consider mixed moments of the type $\bbe(X_{s,t} \cdot X_{s-k,t-l})$. Let $k\geq 1$ \emph{or} $l\geq 1$. Then, the independence assumptions and application of Lemma~1~(i) of \citet{sil_wei_26} yield
\begin{align*}
    \bbe(X_{s,t} \cdot X_{s-k,t-l}) 
    &= \sum_{(i,j)\in\mathcal{S}} \phi_{ij} \cdot \bbe\big((\alpha\circ_{s,t} X_{s-i,t-j}+\varepsilon_{s,t}) \cdot X_{s-k,t-l}\big) \\
    &= \mu_\varepsilon \cdot \mu_X + \sum_{(i,j)\in\mathcal{S}} \phi_{ij} \cdot \bbe\big((\alpha\circ_{s,t} X_{s-i,t-j}) \cdot X_{s-k,t-l}\big) 
    = \mu_\varepsilon \cdot \mu_X + \alpha \sum_{(i,j)\in\mathcal{S}} \phi_{ij} \cdot \bbe(X_{s-i,t-j}\cdot X_{s-k,t-l}),
\end{align*}
because $(s-k,t-l)$ lies in the ``past'' of $(s,t)$. Furthermore, it holds that $\bbe(X_{s-i,t-j}\cdot X_{s-k,t-l}) = \gamma(k-i,l-j) + \mu_X^2$ and $\mu_\varepsilon = (1-\alpha)\mu_X$, so we obtain
\begin{align*}
    \bbe(X_{s,t} \cdot X_{s-k,t-l})
    &= (1-\alpha)\mu_X^2 + \alpha \sum_{(i,j)\in\mathcal{S}} \phi_{ij} \gamma(k-i,l-j) + \alpha \sum_{(i,j)\in\mathcal{S}} \phi_{ij} \mu_X^2 \\
    &= \mu_X^2 + \alpha \sum_{(i,j)\in\mathcal{S}} \phi_{ij} \gamma(k-i,l-j) \qquad \text{for } k\geq 1 \text{ or } l\geq 1.
\end{align*}
Analogously, for $k\leq -1$ \emph{or} $l\leq -1$, we have 
\begin{align*}
    \bbe(X_{s,t} \cdot X_{s-k,t-l})
    &= \mu_\varepsilon\mu_X + \sum_{(i,j)\in\mathcal{S}} \phi_{ij} \cdot \bbe\big(X_{s,t} \cdot (\alpha \circ_{s-k,t-l} X_{s-k-i,t-l-j})\big) \\
    &= (1-\alpha) \mu_X^2 + \alpha \sum_{(i,j)\in\mathcal{S}} \phi_{ij} \big(\gamma(k+i,l+j)+\mu_X^2\big) \\
    &= \mu_X^2 + \alpha \sum_{(i,j)\in\mathcal{S}} \phi_{ij} \gamma(k+i,l+j) \qquad \text{for } k\leq -1 \text{ or } l\leq -1.
\end{align*}
All in all, the proof of Proposition~\ref{prop: CINAR spatial properties} is complete.

\subsection{Proof of Proposition~\ref{prop: CINAR cond distr}}
\label{Proof of Proposition prop: CINAR cond distr}
By the convolution formula and our independence assumptions, we have
\begin{align*}
	\bbp(X_{s,t}=x\, |\,\mathcal{P}\!_{s,t}) &= \bbp\brackets{\sum_{(i,j)\in\mathcal{S}} D_{s,t;i,j} \cdot (\alpha \circ_{s,t} X_{s-i,t-j})+\varepsilon_{s,t}=x\, |\, \mathcal{P}\!_{s,t}} \\
	&= \sum^x_{y=0} \bbp(\varepsilon_{s,t}=y) \cdot \bbp\brackets{\sum_{(i,j)\in\mathcal{S}} D_{s,t;i,j} \cdot (\alpha \circ_{s,t} X_{s-i,t-j})=x-y\, |\, \mathcal{P}\!_{s,t}} \\
	&= \sum^x_{y=0} \bbp(\varepsilon_{s,t}=y) \cdot \sum_{(i,j)\in\mathcal{S}} \phi_{ij} \cdot \bbp(\alpha \circ_{s,t} X_{s-i,t-j}=x-y\, |\, X_{s-i,t-j}).
\end{align*}
The first claim follows from the fact that $\alpha \circ X_{s-i,t-j}|X_{s-i,t-j} \sim \bin(X_{s-i,t-j},\alpha)$ is binomially distributed. Since $\varepsilon_{s,t}$ and $\mathbf{D}_{s,t}$ as well as the thinnings executed at ``time'' $(s,t)$ are independent of the ``past'' $\mathcal{P}\!_{s,t}$, we have
\begin{align*}
	\bbe(X_{s,t}\, |\, \mathcal{P}\!_{s,t}) 
	&= \bbe\, \varepsilon_{s,t} + \sum_{(i,j)\in\mathcal{S}} \phi_{ij}\, \bbe(\alpha\circ_{s,t} X_{s-i,t-j}\, |\, \mathcal{P}\!_{s,t}) \\
	&= \mu_\varepsilon + \sum_{(i,j)\in\mathcal{S}} \phi_{ij}\, \bbe(\alpha\circ_{s,t} X_{s-i,t-j}\, |\, X_{s-i,t-j}) 
    = \mu_\varepsilon + \alpha\sum_{(i,j)\in\mathcal{S}} \phi_{ij}\,X_{s-i,t-j}.
\end{align*}
Concerning the conditional variance, it holds for the model definition in \eqref{eq: CINAR recursion} that
\begin{align*}
	\var(X_{s,t}\, |\, \mathcal{P}\!_{s,t}) 
	&= \sum_{(i,j),(k,l)\in\mathcal{S}} \cov\big(D_{s,t;i,j}\cdot (\alpha \circ_{s,t} X_{s-i,t-j}),\ D_{s,t;k,l}\cdot (\alpha \circ_{s,t} X_{s-k,t-l})\, |\, \mathcal{P}\!_{s,t}\big) \\
	&\hspace{10mm} + 2 \sum_{(i,j)\in\mathcal{S}} \cov\big(D_{s,t;i,j}\cdot (\alpha \circ_{s,t} X_{s-i,t-j}),\ \varepsilon_{s,t}\, |\, \mathcal{P}\!_{s,t}\big) + \var(\varepsilon_{s,t}\, |\, \mathcal{P}\!_{s,t}).
\end{align*}
Due to the independence assumptions on $\varepsilon_{s,t}$, we have $\cov\big(D_{s,t;i,j}\cdot (\alpha \circ_{s,t} X_{s-i,t-j}), \varepsilon_{s,t}\, |\, \mathcal{P}\!_{s,t}\big)=0$ and $\var(\varepsilon_{s,t}\, |\, \mathcal{P}\!_{s,t})=\sigma^2_\varepsilon$, so it remains to consider the conditional covariances appearing in the first line. From the definition of the conditional covariance and the independence assumptions on $\mathbf{D}_{s,t}$, it follows that
\begin{align*}
	\cov\big(D_{s,t;i,j}\cdot (\alpha \circ_{s,t} X_{s-i,t-j}),\, & D_{s,t;k,l}\cdot (\alpha \circ_{s,t} X_{s-k,t-l})\, |\, \mathcal{P}\!_{s,t}\big) \\
	&= \bbe\big(D_{s,t;i,j}\cdot (\alpha \circ_{s,t} X_{s-i,t-j}) \cdot D_{s,t;k,l}\cdot (\alpha \circ_{s,t} X_{s-k,t-l})\, |\, \mathcal{P}\!_{s,t}\big) \\
	&\hspace{10mm} - \bbe\big(D_{s,t;i,j}\cdot (\alpha \circ_{s,t} X_{s-i,t-j})\, |\, \mathcal{P}\!_{s,t}\big)\, \bbe\big(D_{s,t;k,l}\cdot (\alpha \circ_{s,t} X_{s-k,t-l})\, |\, \mathcal{P}\!_{s,t}\big) \\
	&= \bbe(D_{s,t;i,j} \cdot D_{s,t;k,l}) \cdot \bbe\big((\alpha \circ_{s,t} X_{s-i,t-j}) \cdot (\alpha \circ_{s,t} X_{s-k,t-l})\, |\, \mathcal{P}\!_{s,t}\big) \\
	&\hspace{10mm} - \bbe(D_{s,t;i,j}) \cdot \bbe(D_{s,t;k,l}) \cdot \bbe(\alpha \circ_{s,t} X_{s-i,t-j}\, |\, \mathcal{P}\!_{s,t}) \cdot \bbe(\alpha \circ_{s,t} X_{s-k,t-l}\, |\, \mathcal{P}\!_{s,t}).
\end{align*}
As only one component of $\mathbf{D}_{s,t}$ can equal 1 (and the remaining ones equal zero), $\bbe(D_{s,t;i,j} \cdot D_{s,t;k,l})$ equals $\phi_{ij}$ if $(i,j)=(k,l)$ and $0$ otherwise. Therefore, we get
\begin{equation*}
	\cov\big(D_{s,t;i,j}\cdot (\alpha \circ X_{s-i,t-j}),\, D_{s,t;k,l}\cdot (\alpha \circ X_{s-k,t-l})\, |\, \mathcal{P}\!_{s,t}\big)
	= \phi_{ij}\, \bbe\big((\alpha \circ X_{s-i,t-j})^2\, |\, \mathcal{P}\!_{s,t}\big) \cdot \indicator{(i,j)=(k,l)} - \phi_{ij}\phi_{kl}\, \alpha^2\, X_{s-i,t-j} X_{s-k,t-l}.
\end{equation*}
Moreover, it holds that
\begin{equation*}
	\bbe\big((\alpha \circ X_{s-i,t-j})^2\, |\, \mathcal{P}\!_{s,t}\big) = \var(\alpha \circ X_{s-i,t-j}\, |\, \mathcal{P}\!_{s,t}) + \brackets{\bbe(\alpha \circ X_{s-i,t-j}\, |\, \mathcal{P}\!_{s,t})}^2 = \alpha(1-\alpha) X_{s-i,t-j} + \alpha^2 X_{s-i,t-j}^2.
\end{equation*}
Summarizing everything, we obtain the desired result \eqref{eq: CINAR cond var}, and the proof of Proposition~\ref{prop: CINAR cond distr} is complete.

\subsection{Proof of Equations \eqref{eq: tobit CINAR cond dist y}--\eqref{eq: tobit CINAR cond dist x}}
\label{Proof of Equations eq: tobit CINAR cond dist y eq: tobit CINAR cond dist x}
By the convolution formula and our independence assumptions, we have
\begin{align*}
	\bbp(Y_{u,v}=x\, |\,\mathcal{P}\!_{u,v}) &= \bbp\brackets{\sum_{(i,j)\in\mathcal{S}} D_{u,v;i,j} \cdot s(i,j) \cdot (\alpha \circ_{u,v} X_{u-i,v-j})+\varepsilon_{u,v}=x\, |\, \mathcal{P}\!_{u,v}} \\
	&= \sum^\infty_{y=0} \bbp(\varepsilon_{u,v}=y) \cdot \bbp\brackets{\sum_{(i,j)\in\mathcal{S}} D_{u,v;i,j} \cdot s(i,j) \cdot (\alpha \circ_{u,v} X_{u-i,v-j})=x-y\, |\, \mathcal{P}\!_{u,v}} \\
	&= \sum^\infty_{y=0} \bbp(\varepsilon_{u,v}=y) \cdot \sum_{(i,j)\in\mathcal{S}} \phi_{ij} \cdot \bbp\brackets{s(i,j) \cdot (\alpha \circ_{u,v} X_{u-i,v-j})=x-y\, |\, X_{u-i,v-j}} \\
	&= \sum^\infty_{y=0} \bbp(\varepsilon_{u,v}=y) \cdot \sum_{(i,j)\in\mathcal{S}} \phi_{ij} \cdot \bbp\brackets{\alpha \circ_{u,v} X_{u-i,v-j}=s(i,j) \cdot (x-y)\, |\, X_{u-i,v-j}}.
\end{align*}
Here, $\alpha \circ X_{u-i,v-j}|X_{u-i,v-j} \sim \bin(X_{u-i,v-j},\alpha)$ is binomially distributed, i.e., it has non-zero probability masses only if $s(i,j) \cdot (x-y) \in \{0,\ldots,X_{u-i,v-j}\}$ such that the infinite sum $\sum^\infty_{y=0}$ always reduces to a finite one for each $(i,j)\in\mathcal{S}$. 

\smallskip
Also note that $Y_{u,v}$ can only attain negative values if at least one $s(i,j)=-1$. Then, by the model recursion \eqref{eq: tobit CINAR recursion}, the strongest negative value of~$Y_{u,v}$ is attained if all positive summands are equal to zero, and if all negative summands become maximally negative. By the definition of binomial thinning, $\alpha \circ_{u,v} X_{u-i,v-j} \leq X_{u-i,v-j}$, so the strongest negative value of~$Y_{u,v}$ is given by $L_{u,v} := \sum_{(i,j)\in\mathcal{S}} \min\big\{0, s(i,j)\big\}\cdot X_{u-i,v-j}$. 

\smallskip
Finally, due to the relation $X_{u,v} = \max\{0,\ Y_{u,v}\}$, it holds that $X_{u,v}$ is equal to some truly positive value $x>0$ iff $Y_{u,v}=x$. By contrast, any $Y_{u,v}\leq 0$ leads to $X_{u,v} = 0$. So
$$
\bbp(X_{u,v}=0\, |\,\mathcal{P}\!_{u,v}) = \sum_{z=L_{u,v}}^0 \bbp(Y_{u,v}=z\, |\,\mathcal{P}\!_{u,v})
$$
follows, and the proof of Equations \eqref{eq: tobit CINAR cond dist y}--\eqref{eq: tobit CINAR cond dist x} is complete.

\clearpage

\section{Tabulated Simulation Results from Section~\ref{section: Performance of Parameter Estimation}}
\label{appendix: Tabulated Simulation Results}

\begin{table}[bh!]
\centering
\caption{Mean and standard deviation of simulated estimates for Poi-CINAR$(1,1)$ random field, see Section~\ref{section: Performance of Parameter Estimation}.}
\label{tab_sim_poicinar11}

\smallskip
\resizebox{\linewidth}{!}{
\begin{tabular}{r|cccc|cccc|cccc}
\multicolumn{13}{l}{\bf Model\quad $(\mu_\varepsilon,\theta_{01},\theta_{10},\theta_{11}) = (1, 0.1,0.1,0.1)$}\\[1ex]
\toprule
 & $\mu_\varepsilon$ & $\theta_{01}$ & $\theta_{10}$ & $\theta_{11}$ & $\mu_\varepsilon$ & $\theta_{01}$ & $\theta_{10}$ & $\theta_{11}$ & $\mu_\varepsilon$ & $\theta_{01}$ & $\theta_{10}$ & $\theta_{11}$ \\
 \midrule
Mean: & \multicolumn{4}{c|}{YW estimation} & \multicolumn{4}{c|}{CLS estimation} & \multicolumn{4}{c}{CML estimation} \\
\midrule
$(10, 10)$ & 1.067 & 0.086 & 0.085 & 0.080 & 1.006 & 0.096 & 0.097 & 0.101 & 0.999 & 0.097 & 0.099 & 0.102 \\
$(15, 15)$ & 1.058 & 0.090 & 0.087 & 0.081 & 1.019 & 0.097 & 0.095 & 0.094 & 1.017 & 0.098 & 0.094 & 0.095 \\
$(20, 20)$ & 1.041 & 0.091 & 0.092 & 0.088 & 1.014 & 0.097 & 0.098 & 0.098 & 1.010 & 0.097 & 0.099 & 0.098 \\
$(50, 50)$ & 1.011 & 0.098 & 0.098 & 0.096 & 1.000 & 0.100 & 0.100 & 0.100 & 1.000 & 0.100 & 0.100 & 0.100 \\
%$(100, 100)$ & 1.006 & 0.099 & 0.099 & 0.098 & 1.001 & 0.100 & 0.100 & 0.100 & 1.001 & 0.100 & 0.100 & 0.100 \\
\midrule
Std. dev.: & \multicolumn{4}{c|}{YW estimation} & \multicolumn{4}{c|}{CLS estimation} & \multicolumn{4}{c}{CML estimation} \\
\midrule
$(10, 10)$ & 0.203 & 0.081 & 0.081 & 0.075 & 0.230 & 0.093 & 0.092 & 0.095 & 0.219 & 0.093 & 0.094 & 0.093 \\
$(15, 15)$ & 0.150 & 0.060 & 0.061 & 0.057 & 0.165 & 0.066 & 0.066 & 0.066 & 0.153 & 0.065 & 0.065 & 0.066 \\
$(20, 20)$ & 0.124 & 0.049 & 0.050 & 0.048 & 0.132 & 0.052 & 0.053 & 0.053 & 0.124 & 0.051 & 0.052 & 0.052 \\
$(50, 50)$ & 0.048 & 0.020 & 0.021 & 0.021 & 0.049 & 0.020 & 0.021 & 0.021 & 0.043 & 0.019 & 0.020 & 0.020 \\
%$(100, 100)$ & 0.024 & 0.010 & 0.011 & 0.010 & 0.024 & 0.010 & 0.011 & 0.011 & 0.022 & 0.010 & 0.010 & 0.010 \\
\bottomrule
\multicolumn{13}{l}{} \\
\multicolumn{13}{l}{\bf Model\quad $(\mu_\varepsilon,\theta_{01},\theta_{10},\theta_{11}) = (1, 0.2,0.2,0.5)$}\\[1ex]
\toprule
 & $\mu_\varepsilon$ & $\theta_{01}$ & $\theta_{10}$ & $\theta_{11}$ & $\mu_\varepsilon$ & $\theta_{01}$ & $\theta_{10}$ & $\theta_{11}$ & $\mu_\varepsilon$ & $\theta_{01}$ & $\theta_{10}$ & $\theta_{11}$ \\
 \midrule
Mean: & \multicolumn{4}{c|}{YW estimation} & \multicolumn{4}{c|}{CLS estimation} & \multicolumn{4}{c}{CML estimation} \\
\midrule
$(10, 10)$ & 2.393 & 0.214 & 0.214 & 0.332 & 1.584 & 0.185 & 0.182 & 0.476 & 1.015 & 0.199 & 0.200 & 0.500 \\
$(15, 15)$ & 1.809 & 0.217 & 0.217 & 0.384 & 1.313 & 0.188 & 0.190 & 0.490 & 1.009 & 0.198 & 0.199 & 0.502 \\
$(20, 20)$ & 1.555 & 0.215 & 0.216 & 0.413 & 1.218 & 0.191 & 0.192 & 0.494 & 1.004 & 0.198 & 0.199 & 0.502 \\
$(50, 50)$ & 1.168 & 0.211 & 0.209 & 0.464 & 1.042 & 0.199 & 0.197 & 0.499 & 1.004 & 0.201 & 0.199 & 0.500 \\
%$(100, 100)$ & 1.075 & 0.205 & 0.205 & 0.482 & 1.013 & 0.199 & 0.199 & 0.500 & 0.999 & 0.200 & 0.200 & 0.500 \\
\midrule
Std. dev.: & \multicolumn{4}{c|}{YW estimation} & \multicolumn{4}{c|}{CLS estimation} & \multicolumn{4}{c}{CML estimation} \\
\midrule
$(10, 10)$ & 1.069 & 0.096 & 0.103 & 0.107 & 1.094 & 0.102 & 0.107 & 0.125 & 0.254 & 0.087 & 0.085 & 0.104 \\
$(15, 15)$ & 0.714 & 0.069 & 0.070 & 0.076 & 0.728 & 0.069 & 0.071 & 0.082 & 0.157 & 0.049 & 0.052 & 0.060 \\
$(20, 20)$ & 0.482 & 0.053 & 0.052 & 0.060 & 0.485 & 0.054 & 0.054 & 0.062 & 0.119 & 0.040 & 0.038 & 0.046 \\
$(50, 50)$ & 0.182 & 0.020 & 0.021 & 0.024 & 0.183 & 0.021 & 0.021 & 0.025 & 0.044 & 0.014 & 0.014 & 0.018 \\
%$(100, 100)$ & 0.085 & 0.010 & 0.010 & 0.012 & 0.084 & 0.010 & 0.011 & 0.012 & 0.021 & 0.007 & 0.007 & 0.009 \\
\bottomrule
\multicolumn{13}{l}{} \\
\multicolumn{13}{l}{\bf Model\quad $(\mu_\varepsilon,\theta_{01},\theta_{10},\theta_{11}) = (1, 0.3,0.4,0.1)$}\\[1ex]
\toprule
 & $\mu_\varepsilon$ & $\theta_{01}$ & $\theta_{10}$ & $\theta_{11}$ & $\mu_\varepsilon$ & $\theta_{01}$ & $\theta_{10}$ & $\theta_{11}$ & $\mu_\varepsilon$ & $\theta_{01}$ & $\theta_{10}$ & $\theta_{11}$ \\
 \midrule
Mean: & \multicolumn{4}{c|}{YW estimation} & \multicolumn{4}{c|}{CLS estimation} & \multicolumn{4}{c}{CML estimation} \\
\midrule
$(10, 10)$ & 1.592 & 0.261 & 0.345 & 0.074 & 1.333 & 0.260 & 0.357 & 0.114 & 1.017 & 0.293 & 0.393 & 0.110 \\
$(15, 15)$ & 1.348 & 0.285 & 0.369 & 0.074 & 1.176 & 0.282 & 0.378 & 0.103 & 1.013 & 0.297 & 0.397 & 0.102 \\
$(20, 20)$ & 1.228 & 0.294 & 0.383 & 0.077 & 1.090 & 0.291 & 0.391 & 0.099 & 1.009 & 0.298 & 0.399 & 0.100 \\
$(50, 50)$ & 1.074 & 0.300 & 0.395 & 0.090 & 1.017 & 0.299 & 0.398 & 0.100 & 1.002 & 0.300 & 0.399 & 0.100 \\
%$(100, 100)$ & 1.031 & 0.300 & 0.398 & 0.095 & 1.003 & 0.300 & 0.400 & 0.100 & 0.999 & 0.300 & 0.400 & 0.100 \\
\midrule
Std. dev.: & \multicolumn{4}{c|}{YW estimation} & \multicolumn{4}{c|}{CLS estimation} & \multicolumn{4}{c}{CML estimation} \\
\midrule
$(10, 10)$ & 0.627 & 0.106 & 0.103 & 0.078 & 0.656 & 0.115 & 0.115 & 0.106 & 0.253 & 0.095 & 0.095 & 0.094 \\
$(15, 15)$ & 0.391 & 0.073 & 0.071 & 0.061 & 0.406 & 0.076 & 0.078 & 0.074 & 0.159 & 0.063 & 0.058 & 0.062 \\
$(20, 20)$ & 0.276 & 0.055 & 0.053 & 0.051 & 0.279 & 0.059 & 0.057 & 0.059 & 0.119 & 0.044 & 0.045 & 0.048 \\
$(50, 50)$ & 0.108 & 0.023 & 0.023 & 0.024 & 0.109 & 0.024 & 0.023 & 0.025 & 0.045 & 0.018 & 0.018 & 0.018 \\
%$(100, 100)$ & 0.051 & 0.011 & 0.011 & 0.012 & 0.051 & 0.011 & 0.011 & 0.012 & 0.022 & 0.008 & 0.008 & 0.009 \\
\bottomrule
\end{tabular}}
\end{table}

\clearpage

\begin{sidewaystable}[th!]
\centering
\caption{Mean and standard deviation of simulated estimates for NB-CINAR$(1,1)$ random field, see Section~\ref{section: Performance of Parameter Estimation}.}
\label{tab_sim_nbcinar11}

\smallskip
\resizebox{\linewidth}{!}{
\begin{tabular}{r|ccccc|cccc|cccc|ccccc}
\multicolumn{19}{l}{\bf Model\quad $(\mu_\varepsilon, I_\varepsilon,\theta_{01},\theta_{10},\theta_{11}) = (1,2, 0.1,0.1,0.1)$}\\[1ex]
\toprule
 & $\mu_\varepsilon$ & $I_\varepsilon$ & $\theta_{01}$ & $\theta_{10}$ & $\theta_{11}$ & $\mu_\varepsilon$ & $\theta_{01}$ & $\theta_{10}$ & $\theta_{11}$ & $\mu_\varepsilon$ & $\theta_{01}$ & $\theta_{10}$ & $\theta_{11}$ & $\mu_\varepsilon$ & $I_\varepsilon$ & $\theta_{01}$ & $\theta_{10}$ & $\theta_{11}$ \\
 \midrule
Mean: & \multicolumn{5}{c|}{YW estimation} & \multicolumn{4}{c|}{CLS estimation} & \multicolumn{4}{c|}{P-CML estimation} & \multicolumn{5}{c}{N-CML estimation} \\
\midrule
$(10, 10)$ & 1.067 & 1.954 & 0.085 & 0.087 & 0.078 & 1.004 & 0.097 & 0.100 & 0.098 & 1.071 & 0.084 & 0.084 & 0.083 & 1.024 & 1.977 & 0.094 & 0.093 & 0.094 \\
$(15, 15)$ & 1.053 & 1.961 & 0.088 & 0.089 & 0.084 & 1.012 & 0.095 & 0.096 & 0.097 & 1.060 & 0.086 & 0.085 & 0.085 & 1.012 & 1.991 & 0.097 & 0.094 & 0.097 \\
$(20, 20)$ & 1.038 & 1.971 & 0.090 & 0.093 & 0.089 & 1.009 & 0.095 & 0.097 & 0.098 & 1.058 & 0.085 & 0.087 & 0.086 & 1.013 & 1.986 & 0.095 & 0.097 & 0.097 \\
$(50, 50)$ & 1.014 & 1.990 & 0.097 & 0.098 & 0.096 & 1.003 & 0.099 & 0.100 & 0.100 & 1.051 & 0.089 & 0.090 & 0.088 & 1.002 & 1.996 & 0.100 & 0.100 & 0.099 \\
\midrule
Std. dev.: & \multicolumn{5}{c|}{YW estimation} & \multicolumn{4}{c|}{CLS estimation} & \multicolumn{4}{c|}{P-CML estimation} & \multicolumn{5}{c}{N-CML estimation} \\
\midrule
$(10, 10)$ & 0.222 & 0.484 & 0.081 & 0.082 & 0.073 & 0.247 & 0.094 & 0.096 & 0.092 & 0.232 & 0.072 & 0.072 & 0.072 & 0.227 & 0.500 & 0.081 & 0.080 & 0.082 \\
$(15, 15)$ & 0.170 & 0.315 & 0.063 & 0.063 & 0.060 & 0.183 & 0.068 & 0.069 & 0.069 & 0.162 & 0.051 & 0.052 & 0.052 & 0.155 & 0.314 & 0.058 & 0.059 & 0.060 \\
$(20, 20)$ & 0.126 & 0.229 & 0.049 & 0.053 & 0.050 & 0.137 & 0.052 & 0.057 & 0.055 & 0.116 & 0.037 & 0.041 & 0.039 & 0.115 & 0.224 & 0.044 & 0.047 & 0.045 \\
$(50, 50)$ & 0.053 & 0.099 & 0.022 & 0.021 & 0.021 & 0.054 & 0.023 & 0.021 & 0.022 & 0.045 & 0.015 & 0.015 & 0.015 & 0.043 & 0.087 & 0.017 & 0.017 & 0.017 \\
\bottomrule
\multicolumn{13}{l}{} \\
\multicolumn{13}{l}{\bf Model\quad $(\mu_\varepsilon, I_\varepsilon,\theta_{01},\theta_{10},\theta_{11}) = (1,2, 0.2,0.2,0.5)$}\\[1ex]
\toprule
 & $\mu_\varepsilon$ & $I_\varepsilon$ & $\theta_{01}$ & $\theta_{10}$ & $\theta_{11}$ & $\mu_\varepsilon$ & $\theta_{01}$ & $\theta_{10}$ & $\theta_{11}$ & $\mu_\varepsilon$ & $\theta_{01}$ & $\theta_{10}$ & $\theta_{11}$ & $\mu_\varepsilon$ & $I_\varepsilon$ & $\theta_{01}$ & $\theta_{10}$ & $\theta_{11}$ \\
 \midrule
Mean: & \multicolumn{5}{c|}{YW estimation} & \multicolumn{4}{c|}{CLS estimation} & \multicolumn{4}{c|}{P-CML estimation} & \multicolumn{5}{c}{N-CML estimation} \\
\midrule
$(10, 10)$ & 2.395 & 1.676 & 0.215 & 0.208 & 0.333 & 1.562 & 0.182 & 0.178 & 0.479 & 1.353 & 0.190 & 0.188 & 0.483 & 1.048 & 1.907 & 0.197 & 0.195 & 0.503 \\
$(15, 15)$ & 1.806 & 1.830 & 0.220 & 0.217 & 0.381 & 1.314 & 0.191 & 0.188 & 0.488 & 1.368 & 0.193 & 0.193 & 0.475 & 1.029 & 2.006 & 0.200 & 0.199 & 0.497 \\
$(20, 20)$ & 1.555 & 1.845 & 0.216 & 0.215 & 0.413 & 1.218 & 0.192 & 0.191 & 0.495 & 1.338 & 0.191 & 0.192 & 0.480 & 1.011 & 1.993 & 0.198 & 0.198 & 0.503 \\
$(50, 50)$ & 1.160 & 1.966 & 0.212 & 0.210 & 0.463 & 1.036 & 0.200 & 0.198 & 0.498 & 1.328 & 0.195 & 0.194 & 0.477 & 1.005 & 1.997 & 0.200 & 0.200 & 0.499 \\
\midrule
Std. dev.: & \multicolumn{5}{c|}{YW estimation} & \multicolumn{4}{c|}{CLS estimation} & \multicolumn{4}{c|}{P-CML estimation} & \multicolumn{5}{c}{N-CML estimation} \\
\midrule
$(10, 10)$ & 1.128 & 0.899 & 0.106 & 0.102 & 0.110 & 1.148 & 0.111 & 0.104 & 0.128 & 0.413 & 0.082 & 0.080 & 0.095 & 0.296 & 0.634 & 0.083 & 0.083 & 0.097 \\
$(15, 15)$ & 0.708 & 0.764 & 0.071 & 0.071 & 0.081 & 0.737 & 0.073 & 0.074 & 0.087 & 0.277 & 0.050 & 0.051 & 0.059 & 0.175 & 0.436 & 0.051 & 0.051 & 0.061 \\
$(20, 20)$ & 0.484 & 0.618 & 0.051 & 0.053 & 0.061 & 0.503 & 0.051 & 0.053 & 0.064 & 0.200 & 0.035 & 0.036 & 0.042 & 0.124 & 0.306 & 0.036 & 0.037 & 0.043 \\
$(50, 50)$ & 0.178 & 0.315 & 0.021 & 0.021 & 0.026 & 0.177 & 0.021 & 0.021 & 0.026 & 0.078 & 0.014 & 0.013 & 0.016 & 0.047 & 0.115 & 0.014 & 0.013 & 0.017 \\
\bottomrule
%\multicolumn{13}{l}{} \\
\end{tabular}}
\end{sidewaystable}

\clearpage

\begin{sidewaystable}[th!]
\centering\ContinuedFloat
\caption{Mean and standard deviation of simulated estimates for NB-CINAR$(1,1)$ random field, see Section~\ref{section: Performance of Parameter Estimation}.}
%\label{tab_sim_nbcinar11}

\smallskip
\resizebox{\linewidth}{!}{
\begin{tabular}{r|ccccc|cccc|cccc|ccccc}
\multicolumn{13}{l}{\bf Model\quad $(\mu_\varepsilon, I_\varepsilon,\theta_{01},\theta_{10},\theta_{11}) = (1,2, 0.3,0.4,0.1)$}\\[1ex]
\toprule
 & $\mu_\varepsilon$ & $I_\varepsilon$ & $\theta_{01}$ & $\theta_{10}$ & $\theta_{11}$ & $\mu_\varepsilon$ & $\theta_{01}$ & $\theta_{10}$ & $\theta_{11}$ & $\mu_\varepsilon$ & $\theta_{01}$ & $\theta_{10}$ & $\theta_{11}$ & $\mu_\varepsilon$ & $I_\varepsilon$ & $\theta_{01}$ & $\theta_{10}$ & $\theta_{11}$ \\
 \midrule
Mean: & \multicolumn{5}{c|}{YW estimation} & \multicolumn{4}{c|}{CLS estimation} & \multicolumn{4}{c|}{P-CML estimation} & \multicolumn{5}{c}{N-CML estimation} \\
\midrule
$(10, 10)$ & 1.579 & 1.753 & 0.265 & 0.341 & 0.076 & 1.337 & 0.261 & 0.352 & 0.118 & 1.371 & 0.270 & 0.356 & 0.098 & 1.065 & 1.946 & 0.292 & 0.386 & 0.107 \\
$(15, 15)$ & 1.319 & 1.869 & 0.284 & 0.373 & 0.075 & 1.145 & 0.280 & 0.384 & 0.106 & 1.326 & 0.273 & 0.367 & 0.093 & 1.020 & 1.995 & 0.295 & 0.399 & 0.101 \\
$(20, 20)$ & 1.212 & 1.930 & 0.293 & 0.390 & 0.072 & 1.079 & 0.291 & 0.396 & 0.095 & 1.305 & 0.276 & 0.371 & 0.091 & 1.010 & 1.987 & 0.299 & 0.401 & 0.097 \\
$(50, 50)$ & 1.074 & 1.975 & 0.299 & 0.396 & 0.090 & 1.017 & 0.298 & 0.399 & 0.100 & 1.289 & 0.279 & 0.370 & 0.094 & 1.001 & 1.996 & 0.300 & 0.400 & 0.100 \\
\midrule
Std. dev.: & \multicolumn{5}{c|}{YW estimation} & \multicolumn{4}{c|}{CLS estimation} & \multicolumn{4}{c|}{P-CML estimation} & \multicolumn{5}{c}{N-CML estimation} \\
\midrule
$(10, 10)$ & 0.646 & 0.734 & 0.107 & 0.107 & 0.079 & 0.664 & 0.117 & 0.124 & 0.108 & 0.431 & 0.092 & 0.091 & 0.081 & 0.333 & 0.653 & 0.093 & 0.093 & 0.085 \\
$(15, 15)$ & 0.397 & 0.561 & 0.075 & 0.071 & 0.061 & 0.405 & 0.081 & 0.078 & 0.076 & 0.244 & 0.055 & 0.053 & 0.057 & 0.161 & 0.395 & 0.058 & 0.054 & 0.061 \\
$(20, 20)$ & 0.286 & 0.464 & 0.057 & 0.057 & 0.052 & 0.290 & 0.060 & 0.061 & 0.059 & 0.183 & 0.042 & 0.042 & 0.042 & 0.123 & 0.279 & 0.044 & 0.043 & 0.045 \\
$(50, 50)$ & 0.109 & 0.193 & 0.024 & 0.023 & 0.024 & 0.112 & 0.025 & 0.024 & 0.025 & 0.073 & 0.016 & 0.016 & 0.016 & 0.046 & 0.109 & 0.017 & 0.016 & 0.017 \\
\bottomrule
\end{tabular}}
\end{sidewaystable}

\clearpage

\begin{table}[th!]
\centering
\caption{Mean and standard deviation of simulated estimates for Poi-CINAR$(2,2)$ random field, see Section~\ref{section: Performance of Parameter Estimation}.}
\label{tab_sim_poicinar22}

\smallskip
\resizebox{.9\linewidth}{!}{
\begin{tabular}{r|cccc|ccccccccc}
\multicolumn{14}{l}{\bf Model\quad $(\mu_\varepsilon,\theta_{01},\theta_{02},\theta_{10},\theta_{11},\theta_{12},\theta_{20},\theta_{21},\theta_{22}) = (1, 0.1,\ldots,0.1)$}\\[1ex]
\toprule
 & $\mu_\varepsilon$ & $\theta_{01}$ & $\theta_{10}$ & $\theta_{11}$ & $\mu_\varepsilon$ & $\theta_{01}$ & $\theta_{02}$ & $\theta_{10}$ & $\theta_{11}$ & $\theta_{12}$ & $\theta_{20}$ & $\theta_{21}$ & $\theta_{22}$ \\
 \midrule
Mean: & \multicolumn{4}{c|}{CLS-1 estimation} & \multicolumn{9}{c}{CLS-2 estimation} \\
\midrule
$(10, 10)$ & 3.009 & 0.133 & 0.134 & 0.128 & 1.308 & 0.089 & 0.090 & 0.088 & 0.089 & 0.101 & 0.085 & 0.097 & 0.097 \\
$(15, 15)$ & 2.845 & 0.145 & 0.146 & 0.139 & 1.339 & 0.094 & 0.084 & 0.092 & 0.095 & 0.093 & 0.086 & 0.093 & 0.093 \\
$(20, 20)$ & 2.665 & 0.160 & 0.161 & 0.146 & 1.214 & 0.096 & 0.094 & 0.098 & 0.095 & 0.096 & 0.090 & 0.095 & 0.095 \\
$(50, 50)$ & 2.481 & 0.172 & 0.174 & 0.158 & 1.047 & 0.098 & 0.098 & 0.100 & 0.099 & 0.101 & 0.099 & 0.099 & 0.098 \\
\cmidrule{2-14}
 & \multicolumn{4}{c|}{CML-1 estimation} & \multicolumn{9}{c}{CML-2 estimation} \\
\cmidrule{2-14}
$(10, 10)$ & 2.526 & 0.165 & 0.170 & 0.157 & 1.026 & 0.097 & 0.100 & 0.098 & 0.098 & 0.105 & 0.094 & 0.101 & 0.100 \\
$(15, 15)$ & 2.472 & 0.173 & 0.171 & 0.161 & 1.052 & 0.099 & 0.093 & 0.097 & 0.101 & 0.100 & 0.097 & 0.102 & 0.099 \\
$(20, 20)$ & 2.356 & 0.182 & 0.182 & 0.165 & 1.023 & 0.101 & 0.100 & 0.100 & 0.100 & 0.100 & 0.096 & 0.098 & 0.099 \\
$(50, 50)$ & 2.292 & 0.186 & 0.187 & 0.169 & 1.005 & 0.100 & 0.100 & 0.100 & 0.099 & 0.101 & 0.101 & 0.100 & 0.099 \\
\midrule
Std. dev.: & \multicolumn{4}{c|}{CLS-1 estimation} & \multicolumn{9}{c}{CLS-2 estimation} \\
\midrule
$(10, 10)$ & 1.006 & 0.109 & 0.107 & 0.105 & 0.936 & 0.095 & 0.098 & 0.094 & 0.095 & 0.098 & 0.093 & 0.098 & 0.100 \\
$(15, 15)$ & 0.773 & 0.082 & 0.082 & 0.085 & 0.716 & 0.073 & 0.067 & 0.071 & 0.075 & 0.073 & 0.069 & 0.069 & 0.075 \\
$(20, 20)$ & 0.597 & 0.066 & 0.064 & 0.065 & 0.502 & 0.057 & 0.056 & 0.057 & 0.058 & 0.058 & 0.054 & 0.054 & 0.059 \\
$(50, 50)$ & 0.245 & 0.027 & 0.027 & 0.027 & 0.185 & 0.023 & 0.022 & 0.023 & 0.023 & 0.023 & 0.022 & 0.023 & 0.024 \\
\cmidrule{2-14}
 & \multicolumn{4}{c|}{CML-1 estimation} & \multicolumn{9}{c}{CML-2 estimation} \\
\cmidrule{2-14}
$(10, 10)$ & 0.957 & 0.115 & 0.115 & 0.113 & 0.451 & 0.094 & 0.093 & 0.093 & 0.094 & 0.094 & 0.091 & 0.094 & 0.096 \\
$(15, 15)$ & 0.658 & 0.080 & 0.078 & 0.081 & 0.280 & 0.064 & 0.059 & 0.062 & 0.065 & 0.063 & 0.060 & 0.061 & 0.066 \\
$(20, 20)$ & 0.448 & 0.058 & 0.057 & 0.058 & 0.192 & 0.048 & 0.045 & 0.046 & 0.047 & 0.047 & 0.043 & 0.046 & 0.049 \\
$(50, 50)$ & 0.170 & 0.022 & 0.022 & 0.022 & 0.065 & 0.018 & 0.016 & 0.017 & 0.018 & 0.017 & 0.016 & 0.017 & 0.017 \\
\bottomrule
\multicolumn{14}{l}{} \\
\multicolumn{14}{l}{\bf Model\quad $(\mu_\varepsilon,\theta_{01},\theta_{02},\theta_{10},\theta_{11},\theta_{12},\theta_{20},\theta_{21},\theta_{22}) = (1, 0.15,0.1,0.15,0.15,0.05,0.1,0.05,0.1)$}\\[1ex]
\toprule
 & $\mu_\varepsilon$ & $\theta_{01}$ & $\theta_{10}$ & $\theta_{11}$ & $\mu_\varepsilon$ & $\theta_{01}$ & $\theta_{02}$ & $\theta_{10}$ & $\theta_{11}$ & $\theta_{12}$ & $\theta_{20}$ & $\theta_{21}$ & $\theta_{22}$ \\
 \midrule
Mean: & \multicolumn{4}{c|}{CLS-1 estimation} & \multicolumn{9}{c}{CLS-2 estimation} \\
\midrule
$(10, 10)$ & 3.217 & 0.172 & 0.176 & 0.172 & 1.490 & 0.124 & 0.081 & 0.126 & 0.128 & 0.070 & 0.085 & 0.070 & 0.093 \\
$(15, 15)$ & 2.854 & 0.192 & 0.193 & 0.188 & 1.388 & 0.134 & 0.087 & 0.134 & 0.139 & 0.059 & 0.086 & 0.061 & 0.091 \\
$(20, 20)$ & 2.666 & 0.202 & 0.200 & 0.197 & 1.271 & 0.142 & 0.088 & 0.140 & 0.147 & 0.054 & 0.091 & 0.053 & 0.093 \\
$(50, 50)$ & 2.378 & 0.219 & 0.218 & 0.207 & 1.054 & 0.150 & 0.098 & 0.148 & 0.150 & 0.049 & 0.099 & 0.050 & 0.098 \\
\cmidrule{2-14}
 & \multicolumn{4}{c|}{CML-1 estimation} & \multicolumn{9}{c}{CML-2 estimation} \\
\cmidrule{2-14}
$(10, 10)$ & 2.383 & 0.215 & 0.221 & 0.207 & 1.032 & 0.143 & 0.094 & 0.143 & 0.143 & 0.066 & 0.095 & 0.066 & 0.095 \\
$(15, 15)$ & 2.234 & 0.225 & 0.225 & 0.215 & 1.021 & 0.148 & 0.097 & 0.147 & 0.148 & 0.056 & 0.096 & 0.056 & 0.097 \\
$(20, 20)$ & 2.181 & 0.228 & 0.228 & 0.217 & 1.023 & 0.150 & 0.097 & 0.149 & 0.153 & 0.050 & 0.099 & 0.049 & 0.099 \\
$(50, 50)$ & 2.110 & 0.233 & 0.232 & 0.219 & 1.006 & 0.151 & 0.099 & 0.150 & 0.150 & 0.049 & 0.100 & 0.051 & 0.100 \\
\midrule
Std. dev.: & \multicolumn{4}{c|}{CLS-1 estimation} & \multicolumn{9}{c}{CLS-2 estimation} \\
\midrule
$(10, 10)$ & 1.345 & 0.118 & 0.121 & 0.120 & 1.165 & 0.109 & 0.088 & 0.112 & 0.116 & 0.088 & 0.088 & 0.086 & 0.098 \\
$(15, 15)$ & 0.942 & 0.088 & 0.090 & 0.088 & 0.838 & 0.083 & 0.069 & 0.082 & 0.083 & 0.062 & 0.068 & 0.062 & 0.073 \\
$(20, 20)$ & 0.685 & 0.064 & 0.062 & 0.068 & 0.584 & 0.061 & 0.054 & 0.059 & 0.065 & 0.049 & 0.053 & 0.050 & 0.059 \\
$(50, 50)$ & 0.279 & 0.028 & 0.026 & 0.028 & 0.211 & 0.025 & 0.023 & 0.024 & 0.026 & 0.023 & 0.022 & 0.024 & 0.025 \\
\cmidrule{2-14}
 & \multicolumn{4}{c|}{CML-1 estimation} & \multicolumn{9}{c}{CML-2 estimation} \\
\cmidrule{2-14}
$(10, 10)$ & 1.027 & 0.118 & 0.117 & 0.119 & 0.458 & 0.104 & 0.087 & 0.105 & 0.107 & 0.079 & 0.082 & 0.078 & 0.091 \\
$(15, 15)$ & 0.592 & 0.077 & 0.077 & 0.079 & 0.252 & 0.067 & 0.057 & 0.068 & 0.068 & 0.053 & 0.058 & 0.054 & 0.064 \\
$(20, 20)$ & 0.416 & 0.053 & 0.052 & 0.058 & 0.174 & 0.048 & 0.042 & 0.047 & 0.049 & 0.040 & 0.043 & 0.039 & 0.045 \\
$(50, 50)$ & 0.161 & 0.020 & 0.020 & 0.022 & 0.063 & 0.018 & 0.016 & 0.017 & 0.018 & 0.016 & 0.016 & 0.016 & 0.017 \\
\bottomrule
\end{tabular}}
\end{table}

\end{document}